\documentclass[sigconf,screen]{acmart}
\acmConference[ASE 2022]{The 37th IEEE/ACM International Conference on Automated Software Engineering}{Mon 10 - Fri 14 October 2022}{United States}

\AtBeginDocument{%
  \providecommand\BibTeX{{%
    \normalfont B\kern-0.5em{\scshape i\kern-0.25em b}\kern-0.8em\TeX}}}

\usepackage{calligra}
\usepackage{algorithm}
\usepackage{algorithmicx}
\usepackage{algpseudocode}

\usepackage{xspace}
\usepackage{booktabs}
\usepackage{moreverb}
\usepackage{fontenc}
\usepackage{amsmath}
\usepackage{fancybox}
\usepackage{color}
\usepackage{colortbl}
\usepackage{array}
\usepackage{multirow}
\usepackage{multicol}
\usepackage{listings}
\usepackage{makecell}
\usepackage{graphicx}
\usepackage{setspace}
\usepackage[multiple]{footmisc}
\usepackage{wasysym}

\usepackage{balance}
\usepackage{csvsimple}
\usepackage{longtable}
\usepackage{caption}
\usepackage{url}
\usepackage{lscape}
\usepackage{rotating}
\usepackage{tikz}
\usepackage{enumitem}
\usepackage{subcaption}

\usepackage{marvosym}
\usepackage{pifont}
\usepackage{courier}
\usepackage{afterpage,natbib,lipsum}

\algnewcommand\algorithmicforeach{\textbf{for each}}
\algdef{S}[FOR]{ForEach}[1]{\algorithmicforeach\ #1\ \algorithmicdo}

\newcolumntype{L}[1]{>{\raggedright\let\newline\\\arraybackslash\hspace{0pt}}m{#1}}
\newcolumntype{C}[1]{>{\centering\let\newline\\\arraybackslash\hspace{0pt}}m{#1}}
\newcolumntype{R}[1]{>{\raggedleft\let\newline\\\arraybackslash\hspace{0pt}}m{#1}}

\definecolor{codegreen}{rgb}{0,0.6,0}
\definecolor{codered}{rgb}{1,0,0}
\definecolor{codegray}{rgb}{0.5,0.5,0.5}
\definecolor{codepurple}{rgb}{0.58,0,0.82}
\definecolor{backcolour}{rgb}{0.95,0.95,0.92}
\definecolor{lightgray}{gray}{0.9}

\lstdefinestyle{mystyle}{
    commentstyle=\color{codegreen},
    keywordstyle=\color{magenta},
    numberstyle=\small\color{black},
    stringstyle=\color{codepurple},
    basicstyle=\scriptsize\ttfamily,
    breakatwhitespace=false,
    breaklines=true,
    captionpos=b,
    keepspaces=true,
    showspaces=false,
    showstringspaces=false,
    showtabs=false,
    tabsize=2
}

\lstdefinelanguage{diff}{
  morecomment=[f][\color{blue}]{@@},     
  morecomment=[f][\color{red}]-,         
  morecomment=[f][\color{codegreen}]+,       
  morecomment=[f][\color{red}]{---}, 
  morecomment=[f][\color{codegreen}]{+++}
}

\lstdefinelanguage{text}{
  breaklines=false
}

\setlist{noitemsep} 

\definecolor{darkpastelred}{rgb}{0.76, 0.23, 0.13}
\definecolor{ao(english)}{rgb}{0.0, 0.5, 0.0}

\definecolor{darkpastelred}{rgb}{0.76, 0.23, 0.13}
\definecolor{ao(english)}{rgb}{0.0, 0.5, 0.0}

\definecolor{yellow}{RGB}{255,255,153}
\definecolor{grey}{RGB}{224,224,224}

\newboolean{showcomments}
\setboolean{showcomments}{true}
\ifthenelse{\boolean{showcomments}}
 { \newcommand{\mynote}[2]{
      \fbox{\bfseries\sffamily\scriptsize#1}
        {\small$\blacktriangleright$\textsf{\emph{#2}}$\blacktriangleleft$}}}
        { \newcommand{\mynote}[2]{}}

\definecolor{DarkOrange}{rgb}{0.8,0.3,0.0}
\definecolor{DarkCyan}{rgb}{0.0, 0.55, 0.55}
\definecolor{DarkCyel}{rgb}{1.0, 0.49, 0.0}
\definecolor{yellow-green}{rgb}{0.6, 0.8, 0.2}

\newcolumntype{?}{!{\vrule width 1pt}}

\newcommand{\etal}{\emph{et~al.}\xspace}
\newcommand*{\ie}{i.e., }
\newcommand*{\eg}{e.g., }

\newcommand{\toolname}{\textsc{Quatrain}\xspace}

\newcommand{\defi}[1]{
	\setlength{\fboxrule}{1pt}
	\begin{center}\
		\noindent\fcolorbox{black}{white!10}{
			\begin{minipage}{.92\linewidth}
				#1
			\end{minipage}
		}
	\end{center}
	\smallskip
}

\newcommand{\find}[1]{
	\setlength{\fboxrule}{1pt}
	\begin{center}\
		\noindent\fcolorbox{black}{gray!10}{
		\begin{minipage}{.92\linewidth}
			#1
		\end{minipage}
	}
	\end{center}
	\smallskip
}

\copyrightyear{2022}
\acmYear{2022}
\setcopyright{rightsretained}
\acmConference[ASE '22]{37th IEEE/ACM International Conference on Automated Software Engineering}{October 10--14, 2022}{Rochester, MI, USA}
\acmBooktitle{37th IEEE/ACM International Conference on Automated Software Engineering (ASE '22), October 10--14, 2022, Rochester, MI, USA}
\acmDOI{10.1145/3551349.3556914}
\acmISBN{978-1-4503-9475-8/22/10}

\begin{document}
\title[Is this Change the Answer to that Problem?]{
Is this Change the Answer to that Problem?
}
\subtitle{Correlating Descriptions of Bug and Code Changes for Evaluating Patch Correctness}

\author{Haoye Tian}
\email{haoye.tian@uni.lu}
\affiliation{%
   \institution{University of Luxembourg}
 	\country{Luxembourg}
}

\author{Xunzhu Tang}
\email{xunzhu.tang@uni.lu}
\affiliation{%
  \institution{University of Luxembourg}
 	\country{Luxembourg}
}

\author{Andrew Habib}
\email{andrew.a.habib@gmail.com}
\affiliation{%
  \institution{University of Luxembourg}
  \country{Luxembourg}
}

\author{Shangwen Wang}
\email{wangshangwen13@nudt.edu.cn}
\affiliation{%
  \institution{National University of Defense Technology}
  \country{China}
}

\author{Kui Liu}\authornote{Corresponding author.}
\email{brucekuiliu@gmail.com}
\affiliation{%
  \institution{Huawei}
 	\country{China}
}


\author{Xin Xia}
\email{xin.xia@acm.org}
\affiliation{%
  \institution{Huawei}
  \country{China}
}

\author{Jacques Klein}
\email{jacques.klein@uni.lu}
\affiliation{%
  \institution{University of Luxembourg}
 	\country{Luxembourg}
}

\author{Tegawendé F. Bissyandé}
\email{tegawende.bissyande@uni.lu}
\affiliation{%
  \institution{University of Luxembourg}
 	\country{Luxembourg}
}

\begin{abstract}
Patch correctness has been the focus of automated program repair (APR) in 
recent years due to the propensity of APR tools to generate overfitting 
patches. Given a generated patch, the oracle (e.g., test suites) is generally 
weak in establishing correctness. Therefore, the literature has proposed 
various approaches of leveraging machine learning with engineered and deep 
learned features, or exploring dynamic execution information, to further 
explore the correctness of APR-generated patches. In this work, we propose a 
novel perspective to the problem of patch correctness assessment: {\em a 
correct patch implements changes that ``answer'' to a problem posed by buggy 
behavior.} Concretely, {\bf we turn the patch correctness assessment into a 
Question Answering problem}. 
To tackle this problem, our intuition is that natural language processing can 
provide the necessary representations and models for assessing the semantic 
correlation between a bug (question) and a patch (answer). Specifically, we 
consider as inputs the bug reports as well as the natural language description 
of the generated patches. Our approach, \toolname, first considers 
state-of-the-art commit message generation models to produce the relevant 
inputs associated 
to each generated patch. Then we leverage a neural network architecture to 
learn the semantic correlation between bug reports and commit messages. 
Experiments on a large dataset of 9\,135 patches generated for three bug 
datasets (Defects4j, Bugs.jar and Bears) show that \toolname achieves an AUC of 
0.886 on predicting patch correctness, and recalling 93\% correct patches while 
filtering out 62\% incorrect patches. Our experimental results further 
demonstrate the influence of inputs quality on prediction performance. We 
further perform experiments to highlight that the model indeed learns the 
relationship between bug reports and code change descriptions for the 
prediction. Finally, we compare against prior work and discuss the benefits of 
our approach. 
\end{abstract}


%



\maketitle


\section{Introduction}
\label{sec:intro}

Generate-and-validate techniques have achieved success in automatic program repair (APR) by yielding valid patches for a large number of defects in several benchmarks~\cite{gazzola2017automatic, monperrus2020living,liu2020efficiency,selfapr,10.1145/3510003.3510222, yang2021were}. While such techniques are commonplace, their adoption by the industry faces a critical concern with respect to their practicality: state-of-the-art approaches tend to generate patches that overfit the weak oracles (e.g., test suites)~\cite{long2016analysis,yu2019alleviating,wang2019different,lin2022context}.
Indeed, in practice, patches validated by test cases are only plausible. Most of them will be manually found by practitioners to be incorrect~\cite{liu2020efficiency,durieux2019empirical,liu2021critical}. 

Research on automatic assessment of patch correctness has been prolific in recent years~\cite{tian2020evaluating, tian2022best, wang2020automated, yan2022crex}. We identify mainly two categories leveraging either static or dynamic information.
In the first category, only static information is leveraged to decide on patch correctness. For example, Ye~{\em et~al.}~\cite{ye2021automated} have manually crafted static features of code changes that can be used for training a machine learning (ML) based classifier of patch correctness. Similar approaches based on deep representation learning have been proposed~\cite{tian2020evaluating}. More recently, Tian~{\em et al.}~\cite{tian2022predicting} proposed a system where correctness is decided by checking the static similarity of failing test cases vs the similarity of code changes. 
In the second category, traces of dynamic execution of test suites are 
leveraged for correctness evaluation. To predict patch correctness, Xiong~{\em 
et~al.}~\cite{xiong2018identifying} check the behavioral change of failing test 
case executions. Shariffdeen~{\em et al.}~\cite{shariffdeen2021concolic} relied 
on concolic execution to traverse test inputs and patch spaces to reduce the 
number of patch candidates.

Despite the promising results achieved by the aforementioned approaches to 
patch correctness assessment, we identify one fundamental issue and one 
opportunity that open roads to the new research direction studied in this work. 
As a fundamental issue, we note that state-of-the-art approaches generally 
assess patch correctness by reasoning mostly about the code changes, and 
sometimes also about the test case. However, {\bf the bug itself, which is 
targeted by the generated patch, is seldom explicitly investigated}. Yet, 
patches are written to address a specific buggy behavior. As an opportunity, we 
note that bug reports, while informal, may offer an explicit description of the 
bug, which can be leveraged to assess patch correctness. 

To the best of our knowledge, no prior work has investigated the problem of patch correctness as a question-answering problem. We follow the intuition that when a code base maintainer is presented with a patch, the suggested changes are evaluated with respect to the reported bug. That bug is therefore a question. Bug reports, with their natural language description (cf. Example of Figure~\ref{fig:intro-br}), typically pose the problem. The code changes implementing a patch offer an answer to the problem. The commit message describing these changes (cf. Example of Figure~\ref{fig:intro-cm}) typically presents the solutions to the maintainer. The maintainer can then immediately perceive whether the solution (patch) would be relevant to the problem (bug). This scenario of patch validation by human maintainers may appear naive since there are other aspects that developers consider, including whether the bug is real, whether the changes are riskier, etc. Nevertheless, this constitutes a first screening process that we aim to automate by leveraging recent advances in natural language processing (NLP) and machine learning (ML).

\begin{figure}
    \centering
     \begin{subfigure}[t]{.75\linewidth}
\begin{lstlisting}[language=text,linewidth={\linewidth},frame=tb]
Missing type-checks for var_args notation
\end{lstlisting}
    \vspace{-5pt}
    \caption{Title of the bug report.}
    \label{fig:intro-br}
     \end{subfigure} 
 
 	\vspace{5pt}
     \begin{subfigure}[b]{.75\linewidth}
\begin{lstlisting}[language=text,linewidth={\linewidth},frame=tb]
check var_args properly
\end{lstlisting}
    \vspace{-5pt}
    \caption{The commit message of the developer's patch.}
    \label{fig:intro-cm}
     \end{subfigure}
 	\caption{The bug report of Closure-96 from Defects4J and 
 	the corresponding commit message of the developer's patch.}
\end{figure}


{\bf This paper.} We explore the feasibility of leveraging a deep NLP model to assess the semantic correlation between a bug description and a patch description, towards predicting patch correctness for automated program repair. Our main contributions are as follows:

\begin{itemize}[leftmargin=*]
	\item[\ding{182}] We perform a preliminary validation study to demonstrate 
	that bug and patch descriptions are correlated within a dataset of 
	developer submitted patches. This hypothesis validation constitutes a first 
	finding that opens a novel direction for patch correctness studies using 
	bug artifacts. 
	\item[\ding{183}] We formulate the patch correctness assessment problem as a question answering problem and propose \toolname (Question Answering for Patch Correctness Evaluation), a supervised learning approach that exploits a deep NLP model to classify the relatedness of a bug report with a patch description. 
    \item[\ding{184}] We extensively evaluate the effectiveness of \toolname to 
    identify correct patches as well as filter out incorrect patches among a 
    dataset of 9,135 plausible patches (written by developers or generated by 
    APR tools). Our evaluation further compares \toolname to state-of-the-art 
    dynamic~\cite{xiong2018identifying} and static~\cite{tian2020evaluating} 
    approaches, and demonstrates that \toolname achieves comparable or better 
    performance in terms of AUC, F1, +Recall and -Recall.
	\item[\ding{185}] We conduct an analysis of the impact of inputs quality on the prediction performance. In particular, we show that the software engineering committee could benefit from extended research into the direction of patch summarization (a.k.a. commit message generation). 
\end{itemize}

\noindent
{\bf Availability.} Our artifact, code, and dataset are publicly available at:
\url{https://github.com/Trustworthy-Software/Quatrain}.

The remainder of this paper is organized as follows. Section~\ref{sec:bg} provides information on the related work, presents our intuition and summarizes validation data on the hypothesis of our work. Section~\ref{sec:approach} overviews our proposed approach. Experimental setup and results are then described in Sections~\ref{sec:exp} and \ref{sec:eval} respectively. We provide discussions  in Section~\ref{sec:dis} and conclude in Section~\ref{sec:conc}.

\section{Related Work \& Hypothesis}
\label{sec:bg}
In this section, we describe the related work to highlight the relevance of our work and the novelty of our approach. Then we validate the hypothesis that \toolname builds on. 

\subsection{Related work} 
\noindent
{\bf \em Patch correctness.} Test suites are widely used as the oracle to 
validate the correctness of generated patches in APR. Nonetheless, given that a 
test suite is an imperfect (i.e., non-comprehensive) specification of the 
program correctness, a patch that passes the test suite may still be incorrect: 
such patch is referred to in the literature as an {\em overfitting} patch 
\cite{qi2015analysis,long2016analysis}.
With the increasing interest in program repair, a number of research efforts have been undertaken towards identifying such overfitting patches, aiming at mitigating this limitation in APR~\cite{wang2020automated}. 
In general, existing approaches can be split into two categories based on whether they need to execute test cases: static approaches rely on simple heuristics that are summarized from expert knowledge or extracted with learning-based techniques that capture deep features of the patch. For instance, Tan~\etal~\cite{tan2016anti} enumerated several code change rules that help identify overfitting patches as those that violate them.
Ye~\etal~\cite{ye2021automated} trained a machine learning based classifier, ODS, which is based on manually-crafted features specially designed for assessing patch correctness (e.g., presence of  binary arithmetic operators).
Tian \etal \cite{tian2020evaluating} investigated the feasibility of utilizing representation learning techniques (\eg CC2Vec \cite{hoang2020cc2vec} and code2vec \cite{alon2019code2vec}) for comparing overfitting and correct patches. 
In contrast, the dynamic approaches utilize extra information obtained during test execution, such as the execution results on newly-generated test cases~\cite{ye2019automated2,xin2017identifying,yang2017better}, the test runtime trace \cite{xiong2018identifying}, and the program invariant inferred from the execution \cite{yang2020exploring}.
Specifically, Ye \etal \cite{ye2019automated2} used two test generation tools 
(\ie Evosuite \cite{fraser2011evoSuite} and Randoop \cite{pacheco2007randoop}) 
to generate test cases independent from the original test suite and Xin \etal 
\cite{xin2017identifying} proposed to generate tests that are ad-hoc to cover 
the syntactic differences between the generated patch and the ground-truth (\ie 
developer-provided patch).
A patch is considered as overfitting if it fails any of the newly-generated test cases.
Xiong \etal \cite{xiong2018identifying} on the other hand assumes that the ground-truth is missing. They therefore proposed PATCH-SIM, which builds on the similarities among test execution traces for predicting patch correctness. The basic idea is that a patch is likely to be correct if execution traces of the patched program on previously failing tests are significantly different from those of the buggy program. 

In other directions, Yang \etal \cite{yang2020exploring} proposed a  
heuristic where a patch is considered correct if its inferred invariants 
are identical to those of the ground-truth.
Wang \etal \cite{wang2020automated} performed a systematic study and found that combining dynamic approaches with static features achieves promising results and is thus a potential direction.

Overall, prior works in patch assessment have generally exclusively focused on 
the syntax and semantics of the generated patch, and to some extent of the 
patched program behavior. We observe that the literature has largely overlooked 
the characteristics of the bug itself.
In this work, we introduce a novel perspective of patch assessment where the correlation between the patch and the bug is investigated. 
Our intuition is thus that the description of a correct patch for a specific bug has a latent correlation with that bug's description. 
By leveraging the bug reports as an informal description of the buggy behavior 
and a natural language description of the patch, we propose an NLP-driven 
approach, \toolname, which is capable to discriminate correct patches from 
incorrect ones.

\noindent
{\bf \em Leveraging NLP in program repair.}
Given that the target of program repair is to transform a buggy program into its correct version, a number of recent studies have considered it as a translation task. Building on the software naturalness hypothesis~\cite{hindle2012naturalness}, researchers proposed to apply existing neural machine translation (NMT) techniques generally leveraged in natural language processing. 
Chen \etal \cite{chen2019sequencer} proposed a recurrent neural network (RNN) 
based approach that fixes one-line bugs by translating the buggy line into the 
correct line.
Tufano \etal \cite{tufano2019empirical} designed another RNN based model that works at the method level: the model takes a buggy method as input and generates the entire fixed method as output. 
Lutellier \etal \cite{lutellier2020coconut} proposed to separately encode the buggy line and its surrounding contexts (\ie statements that appear before or after the buggy line).
CURE \cite{jiang2021cure}, a more advanced approach, leverages pre-training techniques to help the model gain knowledge about the rigorous syntax of programming languages and how developers write code.
Another recent study \cite{mashhadi2021applying}  investigated the feasibility of applying a large-scale pre-trained model, CodeBERT, to generate patches.

Overall, while these previous works apply NLP techniques to the patch generation process, our work investigates NLP models for patch correctness assessment.

\noindent
{\bf \em  Leveraging bug reports in software engineering tasks:}
Bug reports are considered as invaluable resources for debugging activities since they typically contain detailed descriptions about the program failures as well as the clues of the fault (usually in the form of stack traces) \cite{bettenburg2008makes}.
A number of studies have exploited bug reports to facilitate diverse software engineering tasks.
Liu \etal \cite{liu2013r2fix} and Koyuncu \etal \cite{koyuncu2019ifixr} investigated the feasibility of building program repair systems  based on bug reports, instead of the traditional test cases. Indeed, the primary motivation of their works is that the required test case in APR for triggering the bug may not be readily available in practice when the bug is reported.
Fazzini \etal \cite{fazzini2018automatically} and Zhao \etal \cite{zhao2019recdroid} explored how to automatically reproduce program failures from bug reports without human intervention.
By leveraging code change patterns mined from bug reports, Khanfir \etal 
\cite{khanfir2020ibir} proposed an approach that injects realistic faults to 
improve mutation testing.
Besides, bug reports have been utilized for constructing high-quality defect benchmarks for software testing \cite{saha2018bugs,kim2021denchmark}.
In our study, we leverage bug reports to model the semantics of the bug and thus better assess the patch correctness.

\subsection{Hypothesis Validation}
\begin{figure}
    \includegraphics[width=0.9\columnwidth]{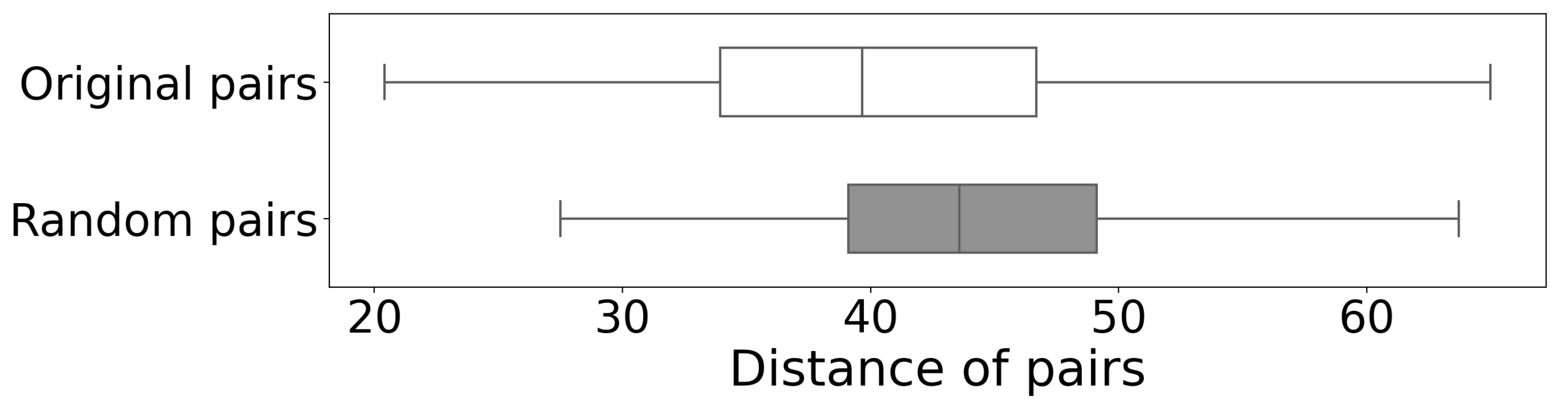}
    \caption{Distributions of Euclidean distances between  bug and patch descriptions.
    }
    \label{fig:euclidean_dis}
\end{figure}

Our hypothesis is that there is a semantic correlation between a bug 
description and the associated (correct) patch description. To validate the 
existence of such a correlation, we conduct a preliminary experiment on a 
collected dataset. The experiment investigates the semantic 
similarity between the descriptions.  To that end, we consider a ground truth 
dataset of Defects4J bugs for which a bug report is available and the commit 
messages describing developer-written patches are provided. These are denoted 
``original pairs''. Then, we assign a randomly selected commit message to each 
bug report in order to build `random pairs'. Finally, to capture the semantics 
of the natural language sentences forming the descriptions (of bugs and 
patches), we utilize a pre-trained deep learning model 
BERT~\cite{devlin2019bert} (introduced in Section~\ref{subsec:gen-vectors}), 
which embeds the descriptions into vectors. We standardize\footnote{In a
standardized dataset, each feature has a range of values with a mean of 0 and 
standard deviation of 1.} these 
vector values to eliminate the influence of dimension on the similarity 
computation. Finally, we calculate Euclidean distance for all pairs. 
Figure~\ref{fig:euclidean_dis} presents the distribution for original pairs and 
random pairs. The results show that the original (\ie ground truth) bug report 
and associated commit message pairs are more similar than random pairs. The 
Mann–Whitney–Wilcoxon test \cite{mcknight2010mann} (p-value: 1.2e-32) further 
validates the significance of the distribution difference. Note that we use 
semantic similarity as a metric to determine correlation. The validation of the 
existence of such correlation motivates the \toolname approach, where NLP 
modelling is leveraged to develop a classification approach of patch 
correctness by building on predicting the relevance of a patch (based on its 
description) for a bug (given its description).

\section{Approach}
\label{sec:approach}
In this section, we first describe the overview of our proposed approach. Then, we fill in the details of the approach with specific steps separated in several subsections. 



\noindent
{\bf [Overview]:}
The intuition we build on is that the natural language description of a bug-fixing patch 
is semantically related to the bug report describing that specific bug: 
the bug report \emph{describes the problem (bug)} and the patch description 
\emph{describes the solution to the problem}.
This semantic relation between a bug report and its associated patch is similar to the QA relation between questions and their answers in NLP.
We present the definition as follow:

\defi{\begin{definition}[Patch Correctness Prediction as a QA Problem]
\label{def:patch-correctness}
Given a bug report in natural language ${br}_{nl}$, a patch ${patch}_{c}$ for 
the reported bug, and a natural language description ${patch}_{nl}$ of the 
patch, predict whether the QA-like pair $({br}_{nl}, {patch}_{nl})$ is
matching or not. 
I.e, predict whether the patch ${patch}_{nl}$ 
is relevant to ({\em answers}) the bug report ${br}_{nl}$ ({\em the question}) or not.
\end{definition}}



\begin{figure*}
    \centering
    \includegraphics[width=0.8\linewidth]{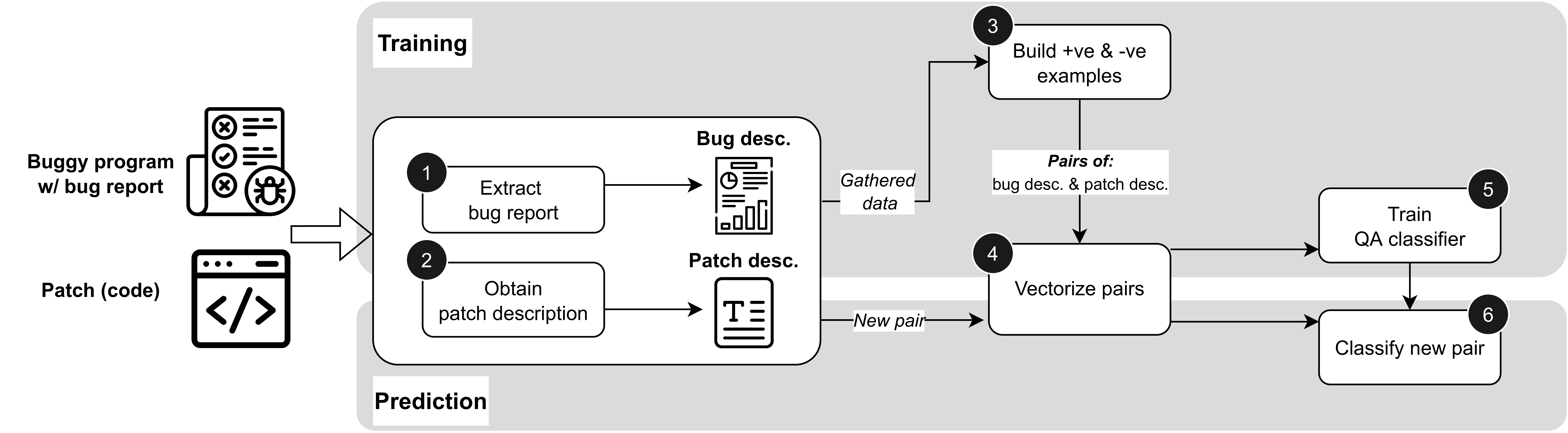}
    \caption{Overview of the approach.}
    \label{fig:overview}
\end{figure*}

To solve this problem, we propose an approach, \toolname, which takes as input 
a program whose buggy behavior is described in a bug report and the associated 
patch generated by an APR tool, and outputs a prediction of correctness. 
Figure~\ref{fig:overview} provides an overview of the approach, which includes 
two phases: training (offline) and prediction (online). 

In the training phase, given a batch of a buggy program, \toolname first 
extracts the bug NL description (bug report) from the program repository in an 
automatic way; then,  for each candidate patch associated to the bug,  it 
generates the patch NL description by leveraging a code change summarization 
tool (e.g., a commit message generator - cf. Sections~\ref{subsec:br} 
and~\ref{subsec:gen-patch}).
Subsequently, \toolname requires a large number of positive (i.e. correct) and 
negative (i.e. incorrect) examples to train a classifier that predicts the 
correlation between a bug report and a patch description. Our third step 
(Section~\ref{subsec:gen-examples}) thus focuses on building a dataset of 
positive and negative examples of pairs of bug reports and associated 
(in)correct patches. In the fourth step (Section~\ref{subsec:gen-vectors}), 
\toolname converts the patch descriptions and the bug reports into vectors in 
high-dimensional vector space to enable model learning. Finally, in the fifth 
step (Section~\ref{subsec:train}), \toolname trains a neural QA classifier on 
the pairs of bug reports and patch descriptions.

In the prediction phase, \toolname pre-processes a new buggy program and its 
associated candidate patch by applying  the first, 
second, and fourth steps in Figure~\ref{fig:overview}. It then uses the trained 
QA classifier to predict whether the candidate patch indeed answers the problem 
in the bug report. The answer is equivalent to a statement on the correctness 
of the 
plausible patch (yes:correct; no:incorrect).

\subsection{Extraction of Bug Reports}
\label{subsec:br}
The first step of our approach is to obtain descriptions of the bugs. 
A natural choice for finding such descriptions is to leverage bug reports. They 
exist in large numbers across projects and provide a NL description of program 
buggy behavior which, at least, describes the symptom of the bug.
Bug reports are submitted via different platforms, e.g., issue trackers such as Jira and issues in GitHub. 
For our purpose, we use a script to automatically mine bug reports for the bug datasets that we use. 

An official bug report typically includes three parts: title, description, and comments.
In the benchmark that we build, some bug reports include comments where the correct solution or even the entire patch is posted. Note, however,  that in our experimental assessment, we must  assume that the bug has not yet been fixed. Thus, to remain practical and reduce bias, we discard all comments and leverage only the title and the description body of bug reports.

\subsection{Generation of Patch Description}
\label{subsec:gen-patch}
The second step of our approach is to summarize an APR-generated patch in 
natural language so as to obtain a semantic explanation of the changes applied in the patch.
The idea here is to get a representation of the patch that is as close as 
possible to how a bug report describes, in natural language, what is the 
bug.
If the patch is written by a developer (e.g., positive example patches in our training set), its associated commit message could be used as a proxy for such NL description of the patch. We use a script to mine the commit messages from developer repositories such as GitHub and collect the descriptions associated to the patches in our datasets. 

Note however that commit messages are obviously not available for APR-generated patches.
Therefore, we automatically generate patch descriptions with the help of state-of-the-art commit message generation techniques. In particular, we consider
CodeTrans~\cite{elnaggar2021codetrans}, an encoder-decoder transformer based model that has been developed to tackle several software engineering tasks.
\toolname uses CodeTrans-TF-Large, the largest such model which achieves the highest BLEU score so far of 44.41 on the commit message generation task.


During training, we obtain patch descriptions either from:
(i) Manually written commit messages of bug-fixing patches provided by developers, or
(ii) Automatically generated descriptions using CodeTrans for APR-generated patches.

\subsection{Construction of Training Examples}
\label{subsec:gen-examples}
Recall from Definition~\ref{def:patch-correctness} that we are addressing a binary classification problem. 
To train a binary classifier, one needs to collect positive (i.e. correct) as well as negative (i.e. incorrect) examples.
Therefore, the third step of \toolname is to build a dataset of positive and negative training examples.

At a high level, a positive (or negative) training (or testing) example consists of a bug report and its associated patch.

\defi{\begin{definition}[Bug report-patch description pair] A {\em bug report-patch description} pair is a tuple $({br}_{nl}, {patch}_{nl})$ of a bug report ${br}_{nl}$ and a patch description ${patch}_{nl}$ (in NL) of a $patch$ that is intended to fix the bug reported in $br$. 
\end{definition}}

\subsubsection{Positive Examples}
We collect two kinds of positive (correct) training examples. 
The first kind of correct examples are developer-written patches and their associated bug reports.
The second kind of correct examples are APR-generated patches and their 
associated bug reports where the APR patches have been manually labeled in 
previous 
studies~\cite{tian2022predicting,ye2021automated,liu2019avatar,li2020dlfix}

\subsubsection{Negative Examples}
We need to create negative examples to train \toolname to identify incorrect patches. 
To do so, we build two kinds of incorrect examples.

For the first kind of negative examples, we randomly assign developer-written patches to irrelevant bug reports. 
For example, we create a training sample by assigning the patch for bug-{\em x} with the bug report of bug-{\em y}. 
The rationale for creating this kind of negative examples is to mobilize the model to learn the hidden relations between bug reports and their associated patch descriptions. 
A patch that tackles a totally irrelevant bug would carry much less - if any - relation to the bug under examination.

The second kind of negative examples is created by selecting APR-generated 
patches that have been labeled as incorrect in previous 
studies~\cite{tian2022predicting,ye2021automated,liu2019avatar,li2020dlfix} and 
their associated bug reports.
The idea is that those APR-generated patches were intended to address the specific bug under examination, but a manual verification revealed that they were incorrect. Correspondingly, the patch description generated for such incorrect patch does not correctly answer the bug report and thus could be considered as negative example.


\subsection{Embedding of Bug Reports and Patches}
\label{subsec:gen-vectors}
To efficiently learn the relationship between bug reports and patch descriptions,  we first need to convert the text into a numerical representations. 
Though there exist various techniques~\cite{bafna2016document,pennington2014glove,church2017word2vec,joulin2016fasttext} for transforming texts into numerical vectors, selecting the proper embedding technique is crucial, as it influences how precisely the vectors represent the text. Compared with popular embedding models such as Word2Vec \cite{mikolov2013efficient}, which uses a fixed representation for each word regardless of the context within which the word appears, BERT~\cite{devlin2019bert} has more advantages for representing texts: it produces word representations that are dynamically informed by the words around them. 
Thus, we employ BERT~\cite{devlin2019bert} as our initial embedding model for both bug reports and patch tokenized texts. The used model is a pre-trained large model
with 24 layers and 1,024 embedding dimension trained on cased English text. 
After representing texts into a vector space, we can perform numerical 
computations on them, e.g., compute text similarity or correlation metrics.

\subsection{Training of the  Neural QA-Model}
\label{subsec:train}
\begin{figure}
    \centering
    \includegraphics[width=0.95\linewidth]{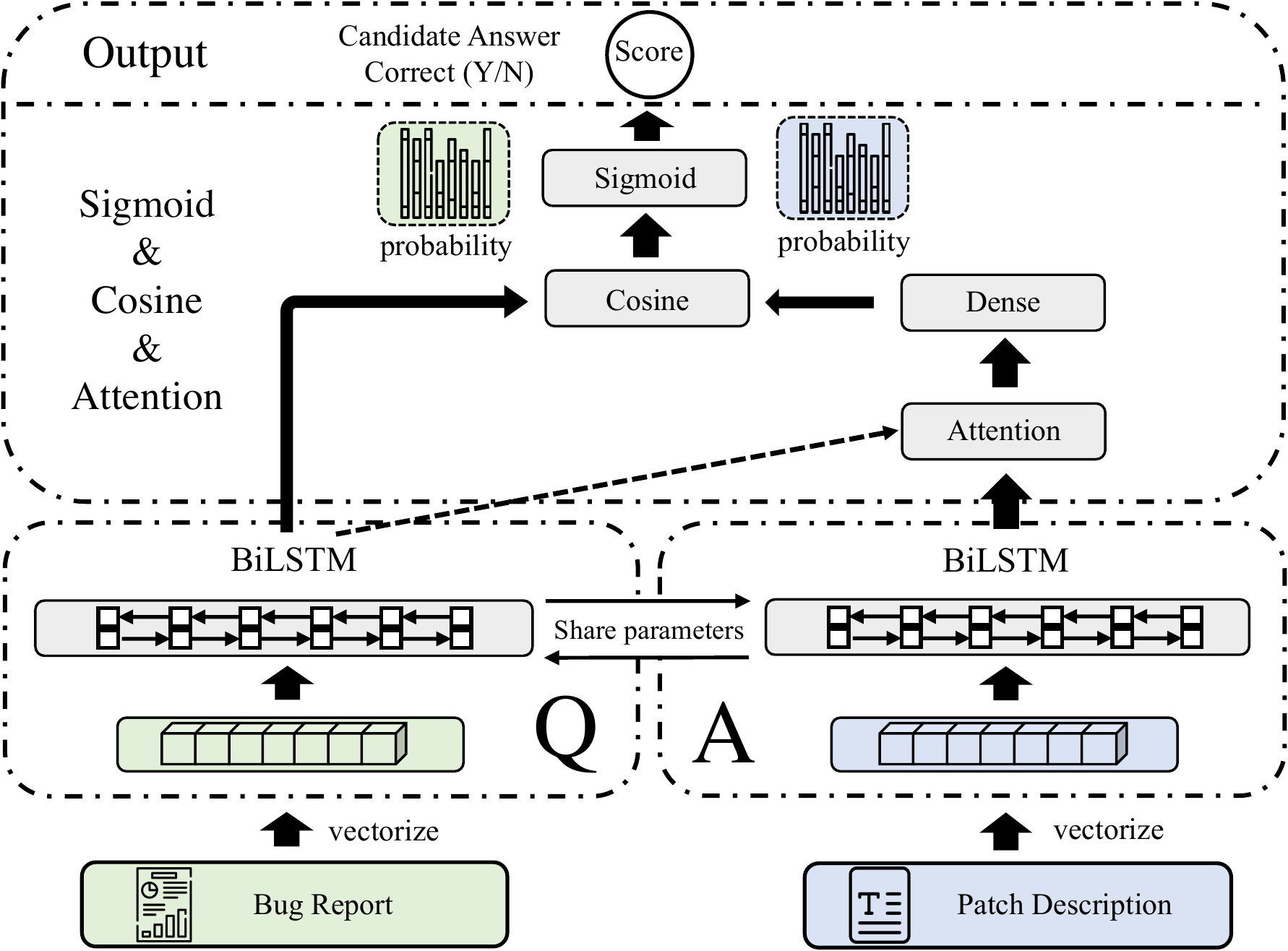}
    \caption{Architecture of the neural QA model.}
    \label{fig:qa-architecture}
\end{figure}

QA-Models are widely applied in Natural Language Processing (NLP) and Software Engineering (SE) communities. Existing literature on QA has addressed various tasks \cite{rajpurkar2016squad, min2017question, gao2021code2que}, among which, the task of answer selection shares fundamental similarities with our bug report-patch description matching problem, \ie select a correct answer (patch) for a question (bug report). Tan \etal \cite{tan2015lstm}s' approach exhibits powerful performance on this task by extending basic bidirectional long short-term memory (BiLSTM) model with an efficient attention mechanism. 
Thus, in the fifth step of \toolname, we propose to adapt the extended QA model from Tan \etal to learn the correlation between bug reports and patch descriptions.
We present the architecture of our adapted QA model in Figure~\ref{fig:qa-architecture}. 

The QA model requires two inputs: bug report and patch description, in their vectorized format (as per  Section~\ref{subsec:gen-vectors}). Then, the BiLSTM layer takes the inputs to learn the correlation between the bug report and the patch description. Assuming input vectors are {\em vector$_b$} and {\em vector$_c$}, we present the BiLSTM in Equation \ref{eq: lstm}.

\begin{equation}
    \begin{split}
        e_b &= BiLSTM(vector_b)= [xb_1, xb_2, ..., xb_N] \in \mathcal{R}^{N 
        \times dim}  \\
        e_c &= BiLSTM(vector_c)= [xc_1, xc_2, ..., xc_N] \in \mathcal{R}^{N 
        \times dim}
    \end{split}
\label{eq: lstm}
\end{equation}
where {\em e$_b$} and {\em e$_c$} represent BiLSTM embeddings of one bug report {\em b} and one associated patch description {\em c}. {\em N} is the length of the input sequence and {\em dim} refers to the dimension size of each sequence. {\em xb$_i$} and {\em xc$_j$} are the embeddings of i-th word in {\em b} and {\em c}.

To better distinguish the correct patch from other patches based on the bug report, we employ an attention mechanism on the patch description to combine the most relevant information according to the bug report, similar to Tan \etal \cite{tan2015lstm}.

To this end, for each word embedding {\em xc$_j$} in patch description, we compute the matrix product $e_b(xc_j)^T$. We then propagate the resulting vector through a softmax operator to obtain the impact weight $\alpha_{xc_j}$ of each word of bug report to {\em xc$_j$}.
\begin{equation}
 \vspace{-0.25cm}
   \begin{split}
        \alpha_{xc_j} = Softmax(e_b(xc_j)^T) \in \mathcal{R}^{N}
    \end{split}
\label{eq: attention}
\end{equation}
where Softmax({\em x}) = $\frac{exp(x)}{\Sigma_iexp(x)}$, and {\em exp(x)} is the element-wise exponentiation of the vector {\em x}. Afterwards, we map the impact weight $\alpha_{xc_j}$ back to each bug report word embedding {\em xb$_i$} to obtain attention representation $att_{xc_j}$,
 \vspace{-0.2cm}



\begin{equation}
 \vspace{-0.2cm}
    \begin{split}
        att_{xc_j} = \Sigma^N_n \alpha_{xc_j,n}xb_i \in \mathcal{R}^{dim}
    \end{split}
\label{eq: weight}
\end{equation}
where $\alpha_{xc_j,n}$ means the {\em n-th} value of $\alpha_{xc_j}$. 
Then, we flatten {\em e$_b$} and $att_{xc_j}$ to one-dimensional vectors {\em re$_b$} and {\em re$_c$} representing the bug report and patch description (with attention), respectively.


Finally, we compute cosine similarity between bug report vector {\em re$_b$} and associated patch description vector {\em re$_c$} and use the sigmoid activation function to normalize the output value of cosine layer to the value range of 0 and 1.
    \begin{equation}\label{eq: sigmoid}
        Score = Sigmoid(cosine(re_b, re_c))
    \end{equation}

where Sigmoid(x) = $\frac{1}{1+exp(-x)}$. The {\em Score} is the prediction probability of patch correctness.







\noindent
{\bf [Hyper-parameters]:}
The employed QA model is mainly based on BiLSTM. We set the max sequence length to 64 and the hidden state dimension size to 16 for the BiLSTM layer. During the training period, we iterate the model parameters by using an Adam optimizer with a leaning rate of 0.01. Considering the data size, we execute 10 training epochs to ensure the convergence of the model. The batch size at each epoch is 128.


\subsection{Classifying a Pair of Bug Report and Patch}
\label{subsec:classify}
For a given buggy program and its APR-generated patch, \toolname classifies the pair as being correlated or not by first extracting the bug description, generating a textual description of the patch, vectorizing the pair of texts, and finally querying the trained QA model. 
A prediction probability is a value between 0 (incorrect) and 1 (correct). 
\toolname labels a patch as being correct or not based on a threshold on the prediction output (Section~\ref{subsec:rq1}).

\section{Experimental Setup}
\label{sec:exp}
We first enumerate the research questions that we investigate to assess the effectiveness of our approach. Then, we describe the dataset used for answering the questions. Finally, we present the evaluation metrics used in our study.

\subsection{Research Questions}
\begin{itemize}[leftmargin=*]
	\item {\em {\bf RQ-1:} What is the effectiveness of \toolname in patch correctness identification based on correlating bug and patch descriptions? } 
	
	We evaluate \toolname on a large dataset consisting of ground truth correct and incorrect patches.
	\item {\em {\bf RQ-2:} To what extent does the quality of the bug report and of the patch description influence the effectiveness of \toolname ? } 
	
	We perform two separate experiments: in the first, we consider the size of texts (\ie number of words) as a proxy for quality, and we investigate whether there is a difference in quality measurement across correct and incorrect predictions. In the second experiment, given the original bug report and developer patch description pairs, we replace them alternatively with a random bug report or a tool-generated patch description and observe changes in performance measurements.
	\item {\em {\bf RQ-3:} How does \toolname perform in comparison with the 
	state-of-the-art techniques for patch correctness identification?} 
	
	We propose to compare our approach against static and dynamic approaches proposed in the literature for APR patch assessment.
\end{itemize}

\subsection{Datasets}

In this paper, we leverage benchmarks that are widely used in the program repair community and on which several APR tools have been applied to generate a large number of patches: Defects4J~\cite{just2014defects4j}, Bugs.jar~\cite{saha2018bugs} and Bears~\cite{madeiral2019bears}. 
Table~\ref{tab:datasets} summarizes the patch dataset that we use for our experiments.
First, we mainly collect the labeled patches (including developer patches) from 
the studies of Tian~{\em et~al.}~\cite{tian2022predicting} and Ye~{\em 
et~al.}~\cite{ye2021automated}. We then supplement the dataset with the patches 
generated by AVATAR~\cite{liu2019avatar} and DLFix~\cite{li2020dlfix}, which 
were not considered in these prior works. Considering that different APR tools 
may generate the same patches for the same bug, we use a simple string-based 
comparison script to deduplicate our patch dataset. Overall, we obtain a large 
duplicated patch assessment dataset of 11,352 patches consisting of 2,260 
correct and 9,092 incorrect patches. Nevertheless, although we removed some 
duplicated patches, there are some semantically equivalent patches that could 
not be detected with our script. For instance, the two conditional statements 
`{\tt if (dataset == null)}` and `{\tt if ((dataset) == null)}` in Java are 
equivalent, although the extra parentheses make their raw strings mismatch. To 
reduce the bias of these duplications in our experiments, we design a specific 
dataset split scheme in Section~\ref{subsec:rq1}.  

\begin{table}
	\centering
	\caption{Datasets of labeled patches.}	
	\resizebox{1\linewidth}{!}{
	\begin{tabular}{l|l|rrr}
			\toprule
		Benchmark & Subjects & Correct & Incorrect &  All \\ 
		\hline
		\multirow{3}{*}{Defects4J~\cite{just2014defects4j}} & Tian~{\em 
		et~al.}~\cite{tian2022predicting} & 1,344 & 1,017 & 2,361 \\
		& Ye~{\em et~al.}\cite{ye2021automated} & 0 & 5,493 & 5,493 \\
		 & AVATAR\cite{liu2019avatar}, DLFix\cite{li2020dlfix} & 59 & 38 & 97 \\
		\hline
		Bugs.jar~\cite{saha2018bugs} & \multirow{2}{*}{Ye~{\em 
		et~al.}\cite{ye2021automated}} & 930 & 2,254 & 3,184 \\
		Bears~\cite{madeiral2019bears} &  & 251 & 531 & 782 \\
		\hline
		{\bf Total} & & 2,584 & 9,333 & 11,917 \\
		\hline
		{\bf Total (deduplicated)} & & 2,260 & 9,092 & 11,352 \\
		\hline
		{\bf Total (experiment)} & & 1,591 & 7,544 & 9,135 \\

		\bottomrule
	\end{tabular}
	}
\label{tab:datasets}
\end{table}


{\bf In our experiment:}
We recall that our approach relies on measuring the correlation between the bug report (BR) and the patch description to predict patch correctness. The collected patches above involve 1,932 unique bugs
To obtain the associated bug reports, we mined their code repositories. Unfortunately, 631 bugs do not contain associated bug reports. Eventually, we were able to leverage 1,301 bug reports. Finally, for the 1,301 unique bugs, we obtain 9,135 available patches consisting of 1,591 correct and 7,544 incorrect patches for our experimental evaluation. 



\subsection{Metrics}
Our objective in patch correctness identification is to recall as many correct 
patches while filtering out as many incorrect patches as possible. 
Thus, we follow the definitions of {\bf Recall} proposed by Tian \etal for the 
evaluation of their BATS~\cite{tian2022predicting}:
\begin{itemize}
	\item {\bf +Recall} measures to what extent correct patches are identified, i.e., the percentage of correct patches that are identified from all correct patches.
	\item {\bf -Recall} measures to what extent incorrect patches are filtered out, i.e., the percentage of incorrect patches that are filtered out from all incorrect patches.
\end{itemize}

\vspace{-8mm}
\begin{multicols}{2}
    \begin{equation}\label{Recall_P}
    + Recall=\frac{TP}{TP+FN}
    \end{equation}
    
    \begin{equation}\label{Recall_N}
    - Recall=\frac{TN}{TN+FP}
    \end{equation}
\end{multicols}

\noindent
where $TP$ represents true positive, $FN$ represents false negative, $FP$ represents false positive, $TN$ represents true negative. 

{\bf Area Under Curve (AUC) and F1}. We construct a deep learning-based NLP classifier to identify the correctness of the patch. Therefore, we use the two most common metrics, AUC and F1 score (harmonic mean between precision and recall for identifying correct patches), to evaluate the overall performance of our approach~\cite{hossin2015review}~. 

\section{Experiments \& Results}
\label{sec:eval}
We conduct several experiments to answer our research questions. In Sections~\ref{subsec:rq1} and~\ref{subsec:rq2}, we focus on the evaluation of approach performance and analysis of approach input to performance. In Section~\ref{subsec:rq3}, 
we compare against the state-of-the-art approaches.

\subsection{Effectiveness of \toolname}
\label{subsec:rq1}

\noindent
{\bf [Experiment Goal]:} We answer RQ-1 by investigating to what extent the \toolname approach, which predicts patch correctness by correlating bug and patch descriptions, is effective. 

\noindent
{\bf [Experiment Design]:} 
In the literature, ML-based approaches to patch correctness identification are 
commonly evaluated using 10-fold cross validation (\ie patch set is divided 
into 90\% for training and 10\% for test)~\cite{tian2020evaluating}. However, 
as we noted in the analysis of our datasets, there are semantically equivalent 
patches. Thus the training and testing set may contain duplicate samples, which 
could lead to biased~\footnote{We discuss this threat in Section~\ref{sec:dis}.} 
experimental results due to data leakage (\ie the model already sees some same 
test samples in the training phase). 

Given the challenge to fully deduplicate the dataset, we propose to limit the 
bias via a new split scheme, referred to as {\em 10-group cross validation}. A 
first manual analysis has shown that the duplicated patches are typically 
generated by different APR tools while targeting the same buggy program. 
Therefore, we first randomly distribute 1,301 unique bugs (including 9,135 
patches) into 10 groups: and every group contains unique bugs and their 
corresponding patches. Then, 9 groups are used as train data and the remaining 
one group is used as the test data. Finally, we repeat the selection of train 
and test groups for ten rounds and  average the metrics obtained across the 
different experimental rounds. 
Through this 10-group cross validation scheme, each patch is able to be leveraged as train data and test data once, which fits the objective of cross-validation. Additionally though, during each train-test process, the unique bugs along with their sets of semantically equivalent patches are exclusively assigned to either train or test group. We trust that such a scheme will provide a realistic evaluation of the performance of learning-based approaches for patch correctness assessment.

\begin{figure}
    \setlength{\abovecaptionskip}{1mm}
	\includegraphics[width=0.9\columnwidth]{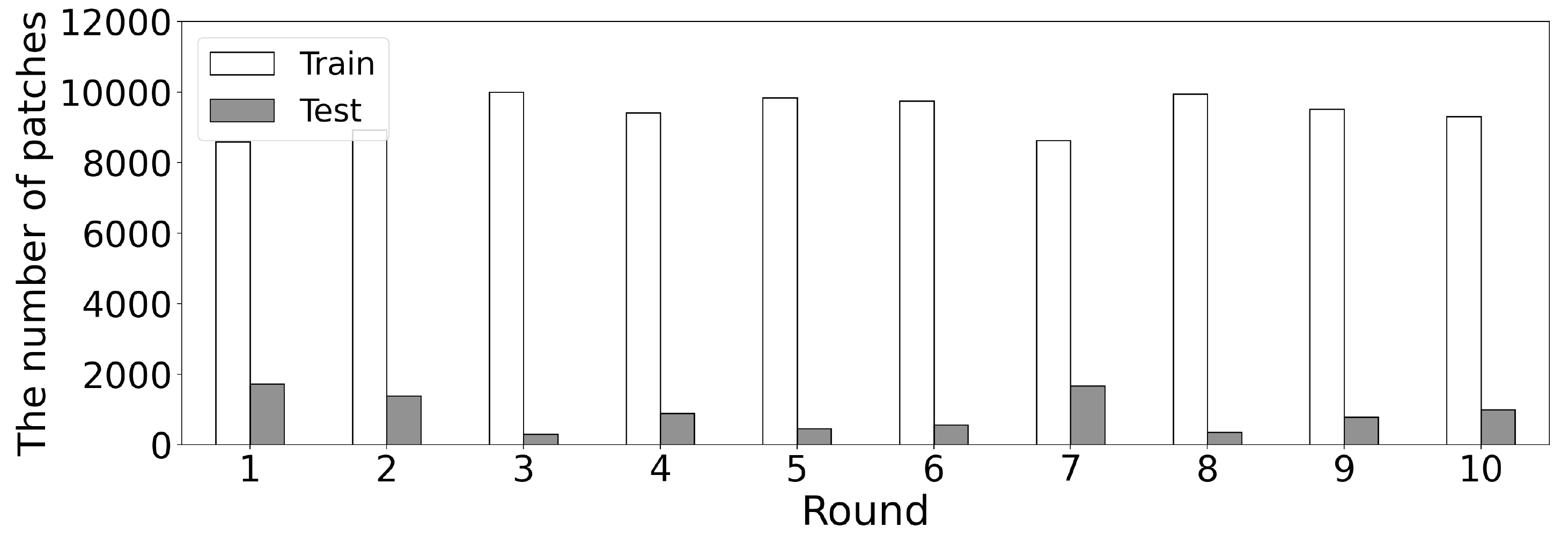}
	\caption{Distribution of Patches in Train and Test Data.}
	\label{fig:data_dis}
	\vspace{1mm}
\end{figure}

Figure~\ref{fig:data_dis} shows the distribution of the number of patches 
assigned to train and test data at each round of 10-group cross validation. The 
overall ratio of train and test data splits is around 10:1. This ratio is close 
to typical 10-fold cross validation (9:1) and thus is appropriate to evaluate 
the performance of train-test based approaches. 

\noindent
{\bf [Experiment Results]:} 
 Using the presented 10-group cross validation, we provide the overall confusion matrix as well as the average +Recall (recall of correct patches) and -Recall (recall of incorrect patches) of \toolname in Table~\ref{tab:matrix}. 
 
 \toolname achieves high AUC at {\bf 0.886}, demonstrating the overall 
 effectiveness of the QA model for patch correctness prediction. We note 
 however that the F1 score (0.628) is relatively low. This metric is known to 
 yield low values when the test data is imbalanced~\cite{jeni2013facing}: in 
 our setting, the ratio is around 5:1 between the incorrect and the correct 
 patch sets. 
 We indeed confirm that that better F1 can be obtained by re-balancing the test 
 data: with a ratio of 1:1 (1,591:1,591) at each round, the same trained 
 classifier achieves a F1 score of {\bf 0.793}.
 Later, in our experiments, we mitigate the potential imbalance bias by 
 comparing against state-of-the-art approaches on the same experimental 
 settings (cf. Section~\ref{subsec:rq3}). 
We found that +Recall and -Recall are sensitive to the selection of thresholds. When setting the threshold at a low value ( \eg 0.1), we are able to identify all correct patches  (+Recall=100\%) but conversely none of the incorrect patches can be filtered out (-Recall=0\%). Similarly, at the threshold value of 0.9, we filter out all incorrect patches but cannot recall any correct patch. Nonetheless, we see that \toolname achieves promising results balanced between +Recall and -Recall when an adequate threshold is selected. 
For instance, \toolname can recall 92.7\% correct patches while filtering out 61.7\% incorrect patches at a threshold value of 0.4
or +Recall of 73.9\% and -Recall of 87.0\% respectively at a threshold of 0.5. 
The results demonstrate our approach is effective on identifying correct and incorrect patches.


\begin{table}
	\centering
	\caption{Confusion matrix of \toolname prediction.}
	\label{tab:matrix}
	\resizebox{1\linewidth}{!}
	{
		\begin{tabular}{c|c|lrrrrrrrrr}
			\toprule
			 \multirow{2}{*}{\bf AUC} & \multirow{2}{*}{\bf F1} & \multicolumn{10}{c}{\bf Thresholds} \\\cline{3-12}
			  & & & 0.1 & 0.2 & 0.3 & 0.4 & 0.5 & 0.6 & 0.7 & 0.8 & 0.9\\
			\hline
			\multirow{6}{*}{0.886} & \multirow{6}{*}{0.628} & \#TP &1,591 &1,582 &1,551 &1,475 &1,175 &583 &189 &0 &0 \\
			 & & \#TN &0 & 2,388& 3,010& 4,653& 6,566& 7,261& 7,522&7,544 &7,544 \\
			 & & \#FP & 7,544& 5,156& 4,534& 2,891& 978&283 &22 &0 &0 \\
			 & & \#FN & 0& 9& 40& 116& 416& 1008& 1,402& 1,591&1,591 \\\cline{3-12}
			 & & +Recall(\%) & 100& 99.4& 97.5& \cellcolor{black!25}92.7& \cellcolor{black!25}73.9& 36.6& 11.9& 0&0 \\
			 & & -Recall(\%) & 0 & 31.7& 39.9& \cellcolor{black!25}61.7& \cellcolor{black!25}87.0& 96.2& 99.7& 100&100 \\
			\bottomrule
		\end{tabular}
	}
\end{table}

\find{{\bf \ding{45} RQ-1} Experimental validation on our collected ground 
truth demonstrates the effectiveness of \toolname in identifying correct 
patches and filtering out incorrect patches: our implementation achieves a 
+Recall of 92.7\% and -Recall of 61.7\% when the decision threshold is set at 
0.4.}

\subsection{The Impact of Input Quality on \toolname}
\label{subsec:rq2}

\noindent
{\bf [Experiment Goal]:} \toolname relies on specific steps to extract bug and patch descriptions once a patch candidate is generated to be applied for a buggy program. The quality of these descriptions may thus influence the performance of our approach. We investigate such an influence by attempting to answer three sub-questions:
\begin{itemize}[leftmargin=*]
	\item {\bf RQ-2.1} {\em To what extent does the length of bug reports and patch descriptions influence the prediction performance?} We hypothesize that good descriptions should have more distinct words, and explore whether correct predictions are made on patch/bug descriptions of larger size.
	\item {\bf RQ-2.2} {\em Does the NLP-based QA classifier actually correlate the bug report and the patch description?} We introduce noise in the test data and evaluate whether the classifier is actually looking at the correlation that we seek to check with the QA.
	\item {\bf RQ-2.3} {\em Do generated patch descriptions provide the same learning value as developer-written commit messages?} We perform experiments where ground truth patch descriptions in the training set are alternatively switched between developer-written (assumed of high quality) and automatically generated (assumed of lower quality) commit messages. 
	\end{itemize}
	
\noindent
{\bf [Experiment Design (RQ-2.1)]:} 
We first define the length of input sentence as the number of distinct words included in the bug reports. Our assumption is that the presence of more distinct words in a textual description may indicate higher quality. Then, for each evaluation round of the 10-group cross validation scheme, we compute the boxplot distribution of length of bug and patch description for correct and incorrect predictions made by our model respectively. Finally, we calculate Mann Whitney Wilcoxon (MWW) to evaluate whether the difference of length is significant across the distributions. The analysis is made on both the length of bug report and patch description. 

\begin{figure}
	\includegraphics[width=0.9\columnwidth]{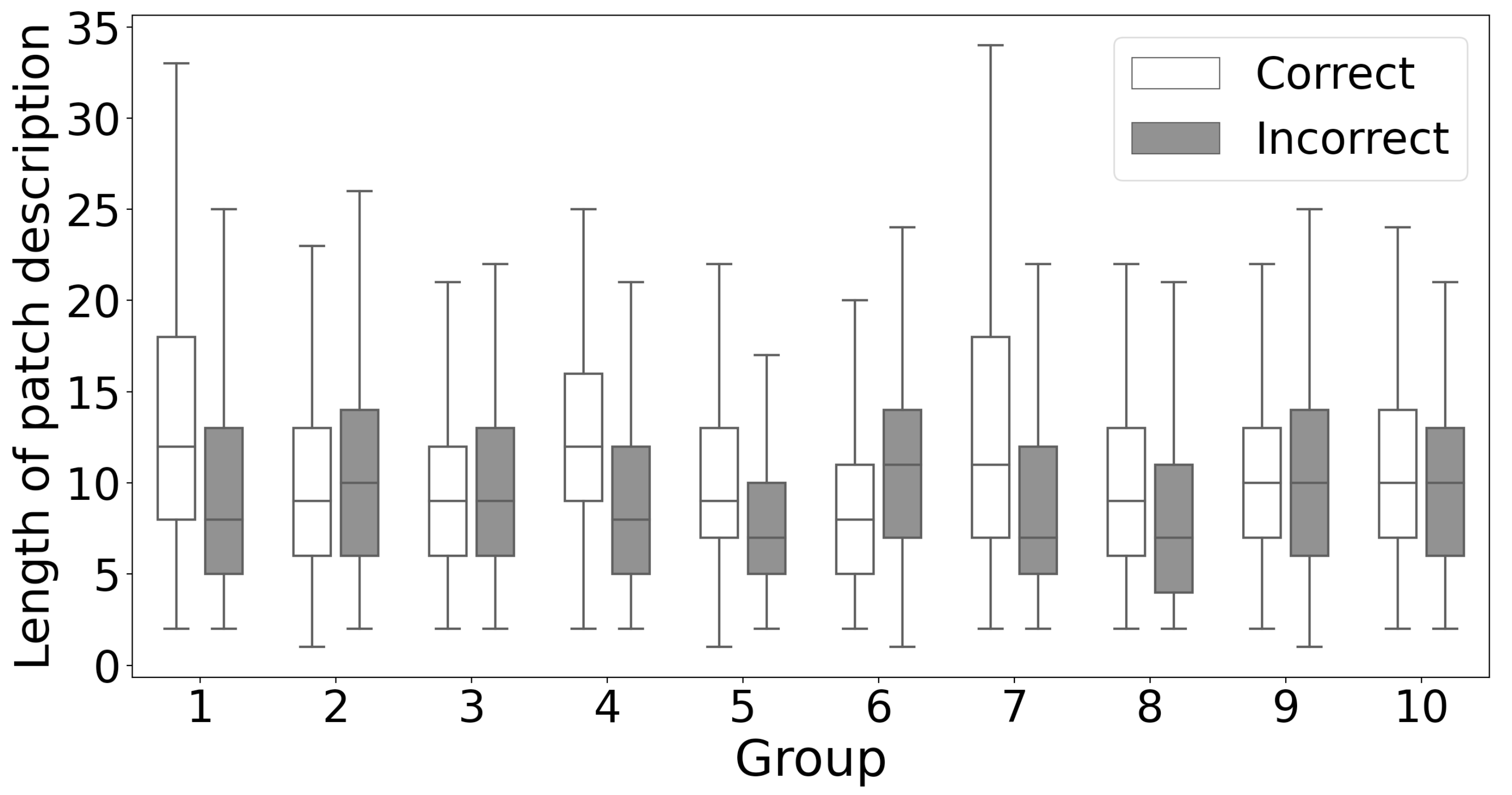}
	\caption{Impact of length of patch description to prediction.}
	\label{fig:lengthCommit}
\end{figure}

\noindent
{\bf [Experiment Results (RQ-2.1)]:} Figure~\ref{fig:lengthCommit} presents the 
distributions of patch description lengths for each round of prediction. We 
observe that, overall in most groups, the length of patch description are 
bigger in the correct predictions than in the incorrect predictions: the model 
is effective when the patch description has larger size.  The 
Mann–Whitney–Wilcoxon test (p-value: 4.1e-16) further confirms that the 
difference of length is statistically significant.
In contrast, the difference for the case of bug reports was not found to be statistically significant.

\find{{\bf \ding{45} RQ-2.1} The higher the quality of patch description (\ie 
in terms of text length), the more \toolname is accurate in predicting patch 
correctness.}

\noindent
{\bf [Experiment Design (RQ-2.2)]:} 
We recall that our NLP model is designed to correlate the bug report and patch description to predict the patch correctness. To validate that some correlation is indeed learned by the devised model, we investigate the influence of associating wrong bug reports to some patches in the test set. We consider the dataset of 1,301 developer-written patches in this experiment since the developer patch description and associated bug report are known to be indeed related by construction. We first compute the performance achieved by \toolname in the prediction of correct patches.
Then, for the patches that \toolname correctly predicts (recall), we re-run the 
classification test where we replace the original bug reports with other 
randomly selected bug reports among the test data. We investigate whether this 
breakdown of the correlation between bug report and patch description is 
reflected in the prediction performance of NLP model.

\begin{figure}
	\includegraphics[width=0.9\columnwidth]{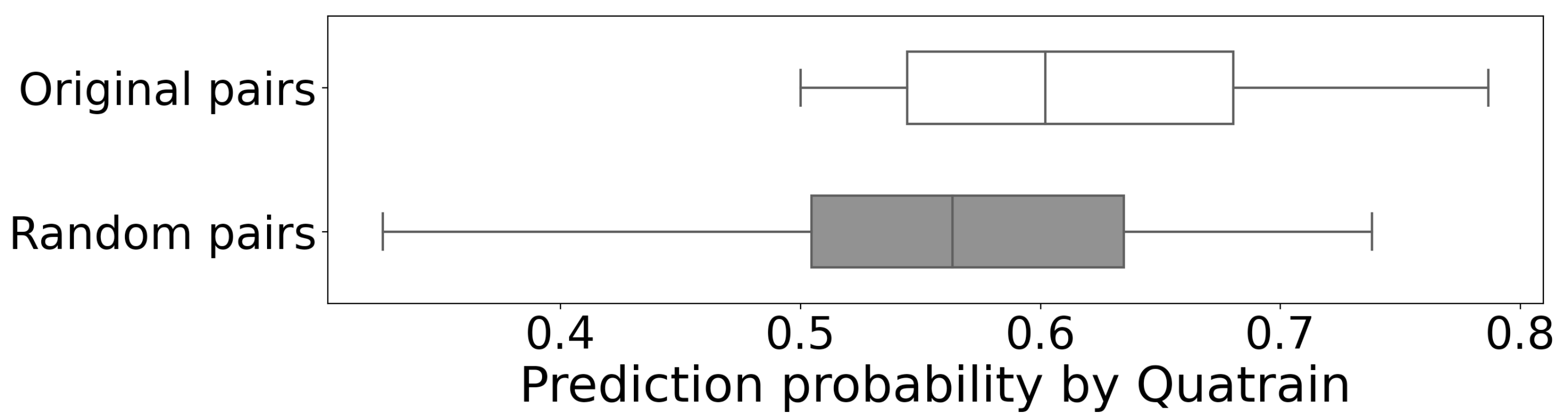}
	\caption{The distribution of probability of patch correctness on original and random bug report.}
	\label{fig:probability}
\end{figure}

\noindent
{\bf [Experiment Results (RQ-2.2)]:} 
Figure~\ref{fig:probability} presents the distribution of prediction 
probability of \toolname for the 1,073 correct patches when the classifier is 
applied on the ground truth pairs (\ie original pairs) and when the classifier 
is applied on pairs where the patch is associated to a random bug report (\ie 
random pairs) . As we see from the boxplot, the lowest value of the 
distribution of original pairs (white box) is around 0.5. This is normal by 
construction: we set 0.5 as the threshold probability for deciding correctness, 
and our data is focused on cases where the prediction was correct. 
After breaking the correlation of bug report and patch description pairs, we found that \toolname yields some prediction probability values smaller than 0.5 (\ie they will be wrongly-classified as incorrect) although the patches are correct.
The Mann–Whitney–Wilcoxon test (p-value:4.0e-35) confirms that the difference 
of median probability values is statistically significant between the two 
distributions. Concretely, 22\% (241/1,073) of developer patches, which were 
previously predicted as correct, are no longer recalled by \toolname after they 
have been associated to a random bug report. These results suggest that  
\toolname indeed assesses the correlation between the bug report and 
the patch description for predicting correctness.

\find{{\bf \ding{45} RQ-2.2} When developer 
patches are paired with random bug reports, \toolname is no longer able to 
predict over 20\% of them correctly. The results suggest that the QA learner in 
\toolname indeed assesses the correlation between the bug report and the patch 
description for predicting correctness.}

\noindent
{\bf [Experiment Design (RQ-2.3)]:}
Commit messages are generally accepted as high-quality descriptions of changes 
since they are manually written by Developers. While CodeTrans is a 
state-of-the-art, its generated-descriptions should be lesser quality. 
Nevertheless, because developer-written commit messages are unavailable in 
practice for APR-generated patches, we must resort to automatic patch 
summarization tools such as CodeTrans.
We evaluate the impact of the quality of patch description (developer-written vs. CodeTrans-generated) on the prediction performance. Our experiments focus on the developer patches only as in RQ-2.2. In the dataset, each patch has two kinds of descriptions, \ie written by developer and generated by Codetrans. We first evaluate our approach based on developer-written descriptions. Then, we replace the developer descriptions with CodeTrans-generated descriptions to assess the performance evolution. 

Besides, we speculate that \toolname is more likely to correctly predict a 
correct patch if the generated description is similar to developer-written 
descriptions used in the training set, we conduct experiments to validate this 
hypothesis. Note however that the semantics of developer-written and generated 
descriptions should be equivalent as they describe the same developer patch. To 
measure the differences in  the descriptions, we adopt the Levenshtein 
distance~\footnote{A classic metric for measuring the distance between two 
strings by calculating the minimal edit operations required. } and compute 
their textual similarity. 

\begin{figure}
	\includegraphics[width=0.9\columnwidth]{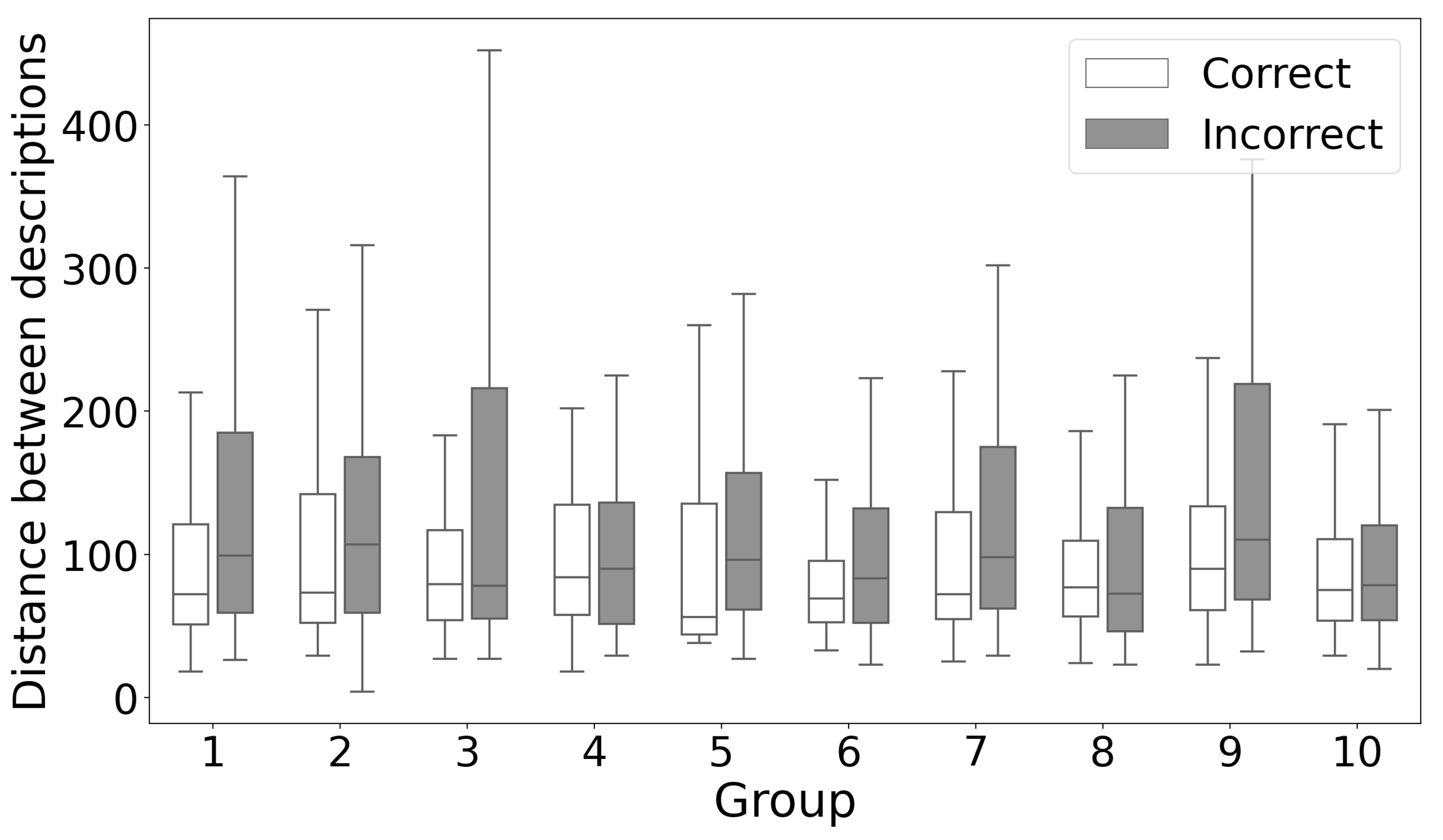}
	\caption{Impact of distance between generated patch description to ground truth on prediction performance}
	\label{fig:distance_lev}
\end{figure}

\noindent
{\bf [Experiment Results (RQ-2.3)]:}
The experimental results show that \toolname achieves a +Recall of 82\% 
(1,073/1,301) when the input for test data uses developer-written descriptions 
of patches as in RQ-2.2. However, that metric (+Recall) drops by 37 percentage 
points to 45\% when the developer-written descriptions are replaced with 
CodeTrans-generated descriptions. This demonstrates that the quality of patch 
description considerably impacts the prediction performance of \toolname.
Figure~\ref{fig:distance_lev} displays the boxplot distribution of Levenshtein distance between developer description and generated description on correct and incorrect predictions respectively. In most of the groups, the white box (correct predictions) presents the shorter Levenshtein distance value, \ie higher similarity. 
This result suggests that if a generated description has a quality that is as high as that of the developer description, \toolname prediction ability will benefit from it. Finally, note that in Section~\ref{subsec:rq1}, we evaluated \toolname in a setting where all developer commit messages were replaced with generated descriptions: the AUC metric dropped by 11 percentage points to 0.774, confirming our findings.

\find{{\bf \ding{45} RQ-2.3 } Patch descriptions generated 
by CodeTrans are often of different quality than ground-truth descriptions. 
Good patch descriptions help \toolname identify more correct patches.}
\subsection{Comparison against the State-of-the-art}
\label{subsec:rq3}

While previous RQs have shown that \toolname is 
effective, in this section we compare it against state-of-the-art static and 
dynamic approaches.
Finally, we investigate the complementarity of \toolname to other existing 
approaches.


\subsubsection{\bf Comparing against Static Approaches}
We compare \toolname against two state-of-the-art approaches: (i) A pure 
classification approach based on patch embeddings~\cite{tian2020evaluating} and 
(ii) BATS which leverages the embedding of test cases to compute similarity 
among failing test cases and among associated patches~\cite{tian2022predicting}.

\subsubsection*{\bf \toolname vs. (supervised) DL-based Patch Classifier}
In \toolname, we first leverage pre-trained Bert model to embed the natural language text of bug report and patch description of patch. Then, we build a deep learning classifier to capture the QA relationship between these descriptions to predict patch correctness. Since Tian~{\em et al.}'s approach also use BERT and construct a classifier for patch correctness validation, we compare our approach against theirs. For a fair comparison, we reproduce their evaluation on our dataset. Concretely, when we train or test our model with divided-by-group patches, we consistently use the same patches for the training and testing of Tian~{\em et al.}'s classifiers of Logistic Regression
(LR) and Random Forrest (RF), following their experimental setup.

Table~\ref{tab:compare_ASE}~presents the comparison results: Tian~{\em et 
al.}'s best classifier (RF) achieves +Recall of 89.4\% while filtering out 
59.8\% incorrect patches. Meanwhile, \toolname achieves a better +Recall of 
92.7\%, and filters out slightly more incorrect patches (-Recall of 61.7\%). 
Regarding the overall performance metrics AUC and F1, \toolname outperforms the 
approach of Tian \etal. We finally investigate the complementarity of our 
approach. Among 9,135 patches, our approach identifies 7,842 patches, of which 
2,735 patches cannot be identified by Tian~{\em et al.}'s approach (RF).

\begin{table}
	\centering
	\caption{\toolname vs a DL-based patch classifier~\cite{tian2020evaluating}.}
	\label{tab:compare_ASE}
	\resizebox{1\linewidth}{!}
	{
		\begin{tabular}{lr|rrrr}
			\toprule
            {\bf Classifier} & {\bf Incorrect:Correct} & {\bf AUC} & {\bf F1} & {\bf +Recall} & {\bf -Recall} \\
			\midrule
			\multirow{1}{*}{Tian et al. (LR)} & 7,544:1,591 (5:1)  & 0.719 & 
			0.449 & 0.833 & 0.605  \\
			\multirow{1}{*}{Tian et al. (RF)} & 7,544:1,591 (5:1)  & 0.746 & 
			0.470 & 0.894 & 0.598  \\
			\midrule
			\multirow{1}{*}{\toolname} & 7,544:1,591 (5:1) & 
			\cellcolor{black!25}0.886 & \cellcolor{black!25}0.628 &  
			\cellcolor{black!25}0.927 & \cellcolor{black!25}0.617 \\
		   \bottomrule
		\end{tabular}
	}
\end{table}

\subsubsection*{\bf \toolname vs. (unsupervised) BATS}
BATS\cite{tian2022predicting} is the most recent patch correctness assessment approach proposed by Tian~{\em et al.}. It is devised based on a simple but novel hypothesis that when different defective programs fail to pass similar test cases, it's likely that the programs can be repaired by similar code changes. Given a buggy program, failing test cases and a plausible patch, BATS first searches the most similar failing test cases from other oracle programs. Afterwards, the associated correct patches that fix these similar test cases are extracted to compute their similarity with the generated plausible patch. BATS labels the plausible patch as correct if that similarity is beyond an inferred threshold, otherwise it is predicted as incorrect. 

According to the authors' open-source artifacts, BATS is currently able to be evaluated on Defects4J and is not adapted for Bears and Bugs.jar. We thus conduct the comparison on the benchmark of Defects4J. Although Tian~{\em et al.} demonstrated that BATS shows promising results on identifying patch correctness, its scalability is limited due to the lack of enough test cases in the search space to compute similarity.
They thus added a cut-off on the similarity computation of test cases to focus on a subset of patches where BATS is applicable. We follow their experimental setup to reproduce BATS evaluation on our dataset with the cut-off of 0.0 (non-specific scenario) and 0.8 (the best performance in their evaluation). We compare our approach on the same available dataset.

As shown in Table~\ref{tab:compare_BATS}, the configuration of cut-off incurs the reduction of patch set that can be evaluated. Our approach comprehensively outperforms BATS whether they filter dissimilar programs or not (cut-off: 0.0 and 0.8). 
Note that BATS is not able to scale its performance, in this scenario, to the entire dataset due to the lack of similar test cases. In addition, in this scenario, 180 out of 345 patches are exclusively identified by \toolname.

\begin{table}
	\centering
	\caption{\toolname vs BATS~\cite{tian2022predicting}.}
	\label{tab:compare_BATS}
	\resizebox{1\linewidth}{!}
	{
		\begin{tabular}{lr|rrrr}
			\toprule
            {\bf Classifier} & {\bf Incorrect:Correct} & {\bf AUC} & {\bf F1} & {\bf +Recall} & {\bf -Recall} \\
			\midrule
			\multirow{1}{*}{BATS (cut-off: 0.0)} & 4,930:385 (13:1)  & 0.549 & 
			0.149  & 0.647 & 0.452  \\
			\multirow{1}{*}{\toolname} & 4,930:385 (13:1) & 
			\cellcolor{black!25}0.824 & 0.350 & \cellcolor{black!25}0.803 & 
			\cellcolor{black!25}0.662 \\
			\midrule
            \multirow{1}{*}{BATS (cut-off: 0.8)} & 367:41 (9:1) & 0.620 & 0.235  & 0.805 & 0.436  \\
			\multirow{1}{*}{\toolname} & 367:41 (9:1) & \cellcolor{black!25}0.832 & 0.462 & \cellcolor{black!25}0.902 & 
			\cellcolor{black!25}0.453  \\
		   \bottomrule
		\end{tabular}
	}
\end{table}

\subsubsection{\bf Comparing against a Dynamic Approach}
We consider a dynamic approach where execution traces are also leveraged in the 
prediction of correctness. PATCH-SIM~\cite{xiong2018identifying} is a state-of-
the-art tool for dynamic assessment of patch correctness: it compares test 
execution information before and after patching a buggy program. The hypothesis 
they proposed is that correct patches tend to change the behavior of execution 
of failing test cases and retain the behavior of passing test cases. 
Due to the failure of prediction for part of the patches~\footnote{We reported 
the problem to the PATCH-SIM authors and we are still waiting for their 
response.} and limitation of timeout, we can apply PATCH-SIM to 3,546 
patches. 
The results in Table~\ref{tab:compare_PATCHSIM} show that our approach filters out more incorrect patches while reaching same +Recall compared to PATCH-SIM. 
Most of the patches (1,856/3,149) that we identify are not correctly predicted 
by PATCH-SIM. Note that the low values of F1 score (both for PATCH-SIM and 
\toolname) are due to the extremely imbalanced ratio of 44:1 in 
incorrect:correct sets.

\begin{table}
	\centering
	\caption{\toolname vs (execution-based) PATCH-SIM~\cite{xiong2018identifying}.}
	\label{tab:compare_PATCHSIM}
	\resizebox{1\linewidth}{!}
	{
		\begin{tabular}{lr|rrrr}
			\toprule
            {\bf Classifier} & {\bf Incorrect:Correct} & {\bf AUC} & {\bf F1} & {\bf +Recall} & {\bf -Recall} \\
			\midrule
			\multirow{1}{*}{PATCH-SIM} & 3,468:78 (44:1) & 0.581 & 0.053 & 
			0.769 & 0.392  \\
			\multirow{1}{*}{\toolname} & 3,468:78 (44:1) & 
			\cellcolor{black!25}0.792 & 0.127 & \cellcolor{black!25}0.769 & 
			\cellcolor{black!25}0.667  \\
		   \bottomrule
		\end{tabular}
	}
\end{table}

\find{{\bf \ding{45} RQ-3} Comparing against state-of-the-art static and 
dynamic approaches, \toolname achieves competitive (or better) 
performance in predicting patch correctness.}

\section{Discussion}
\label{sec:dis}
We enumerate a few insights from our results and discuss the threats to the validity of our study. 
\subsection{Experimental Insights}
\label{subsec: insights}
[{\em Insufficient deduplication of semantically-equivalent patches may lead to biased prediction performance.}] As we mentioned in the experimental design in Section~\ref{subsec:rq1},
the traditional 10-fold cross validation scheme may assign the same semantically-equivalent patches simultaneously into both train and test datasets. 
In practice, this setup violates the principles in machine/deep learning-based 
evaluations since it's equivalent to letting the models cheat by learning 
knowledge from test data during the training process\cite{allamanis2019adverse, 
irolla2018duplication, zhao2021impact}. To showcase this bias in the results, 
we propose to focus on a straightforward classifier using a random forest on 
the embeddings of the bug report and the patch: when using 10-fold cross 
validation scheme on our ground truth dataset, the achieved AUC is as high as 
0.978 (with F1 at 0.860); however, when using our deduplication scheme 
(10-group cross validation based on bug ID), the AUC drops to 0.780 (and F1 at 
0.344).

\noindent
[{\em Generating high quality code change description can help identify patch correctness}] We found that the quality of code change description influences the prediction performance of \toolname. In RQ-2.1 and RQ-2.3, the experimental results show the model makes more correct predictions when addressing longer or more developer written-similar code change description. Our experiments offer some evidence 
to encourage the community to design advanced patch summarization approaches. 
\toolname indeed can become a prime candidate for leveraging such research 
output to further increase the practicality and adoption of automated program 
repair. 

\subsection{Case Study}
Figure~\ref{fig:lang7} presents an example correct patch generated by DLFix, an APR tool, for Defects4J bug Lang-7. \toolname successfully predicts its correctness while BATS fails to do so.
The associated bug~\footnote{https://issues.apache.org/jira/browse/LANG-822} is 
reported as follows:

{\small
\begin{verbatim}
Title: NumberUtils#createNumber - bad behaviour 
for leading "--". 
Description: NumberUtils#createNumber checks for 
a leading "--" in the string, and returns null if 
found. This is documented ...  	
\end{verbatim}
}

\begin{figure}
    \centering
    \scriptsize
\begin{lstlisting}[language=Diff,linewidth={\linewidth},frame=tb]
--- ./src/main/.../NumberUtils.java
+++ ./src/main/.../NumberUtils.java
@@ -449,9 +449,7 @@
         if (StringUtils.isBlank(str)) {
             throw new NumberFormatException("A blank string is not a valid number");
         }  
-        if (str.startsWith("--")) {
-            return null;
-        }
+
         if (str.startsWith("0x") || str.startsWith("-0x") || str.startsWith("0X") || str.startsWith("-0X")) {
\end{lstlisting}
    \caption{A correct generated patch for Defects4J Lang-7.}
    \label{fig:lang7}
\end{figure}

BATS assumes that similar buggy programs require similar patches to fix.
To predict the generated patch correctness, BATS searches for a buggy program 
that fails on similar failing test cases with Lang-7. The retrieved program is 
the bug Lang-16~\footnote{An upper-case hex bug in 
https://issues.apache.org/jira/browse/LANG-746} in Defects4J. BATS predicts the 
generated patch is correct if it's similar with the developer patch addressing 
bug Lang-16. However, the retrieved bug Lang-16 is not related with bug Lang-7 
even though they have similar test cases and require a dissimilar patch to fix. 
Thus, BATS fail to predict the generated patch correctness.

Consider however the NL description of the patch as it is generated by 
CodeTrans:

{\small
\begin{verbatim}
removed the unnecessary "" -- "" from NumberUtils . 
startsWith ( ) , it was restricting our.
\end{verbatim}
}

\smallskip
\noindent The syntactic and semantic correlation between the bug and patch 
description is obvious, which supports the fact that \toolname predicts the 
patch as correct.

\subsection{Threats to Validity}
The implementation of \toolname uses a pre-trained BERT to embed bug and patch 
descriptions before feeding them into the QA model. \toolname also uses 
CodeTrans to generate patch descriptions. These choices may have influenced 
greatly our results. The associated threat is nevertheless limited since these 
constitute the state-of-the-art in their respective domains.

 Our evaluation dataset includes 9,135 patches, though it is highly 
 imbalanced (83\% incorrect vs. 17\% correct patches). This imbalance 
 may bias our results. We mitigate this bias by stressing more 
 on AUC metric, rather than F1 score and by performing comparison experiments 
 against the state-of-the-art.
 
 Our patch correctness labels have been manually decided in prior work~\cite{tian2022predicting}. 
 The accuracy of the labels and the ground truth constitute a threat to 
 validity~\cite{yang2021ground}, which is mitigated in part by our comparison 
 against the state-of-the-art on the same datasets.
 
 Our experimental evaluation does not perform any fine-tuning of the hyper-parameters of the QA model or even the initial BERT model used for embedding bug and patch descriptions. The yielded performance may thus not be representative of what can be achieved. 


\noindent
\section{Conclusion}
\label{sec:conc}
In this paper, we present a novel perspective to the patch correctness 
assessment problem in automated program repair. Given a plausible patch, which 
is validated by an imperfect oracle, the need for correctness identification is 
acute, as several studies have revealed that state-of-the-art repair tools 
generate overfitting patches. 
Our idea is that a correct patch is the one that answers to the problem revealed by the execution failure (bug).
We therefore design \toolname, a neural network architecture that leverages NLP 
to learn to correlate bugs and patch descriptions and produce a 
Question-Answering based classifier. Given a buggy program, we consider its bug 
report and leverage CodeTrans to generate descriptions for all APR-generated 
patches targeting the bug. Then, we use these NL descriptions of bugs and 
patches to feed the QA classifier of \toolname. The classification decision 
serves as a prediction of patch correctness. The experimental results show that 
our approach identifies 92.7\% correct patches and filter 61.7\% incorrect 
patches with an AUC of 0.886. We then investigate and discuss the influence of 
the quality of the input (bug report and code change description) on the 
effectiveness of \toolname. We also perform experiments to demonstrate that 
\toolname indeed learns and builds on the correlation between the bug report 
and the patch to make the predictions. Finally, we reproduce recent 
state-of-the-art static and dynamic patch assessment tools on our dataset and 
show that \toolname exhibits comparable or better effectiveness in recalling 
correct patches and filtering out incorrect patches. Insights from our work 
open new research direction in patch assessment, but also provide a novel use 
case for a large body of the literature that is focused on code summarization.

\begin{acks}
This work was supported by the NATURAL project, which has received funding from 
the European Research Council under the European Union’s Horizon 2020 research 
and innovation program (grant No. 949014).
Kui Liu was also supported by the National Natural Science Foundation of China (Grant No. 62172214), the Natural Science Foundation of Jiangsu Province, China (Grant No. BK20210279), and the Open Project Program of the State Key Laboratory of Mathematical Engineering and Advanced Computing (No. 2020A06).
%
\end{acks}

\balance
\bibliographystyle{ACM-Reference-Format}
\bibliography{references}


\begin{thebibliography}{66}


\ifx \showCODEN    \undefined \def \showCODEN     #1{\unskip}     \fi
\ifx \showDOI      \undefined \def \showDOI       #1{#1}\fi
\ifx \showISBNx    \undefined \def \showISBNx     #1{\unskip}     \fi
\ifx \showISBNxiii \undefined \def \showISBNxiii  #1{\unskip}     \fi
\ifx \showISSN     \undefined \def \showISSN      #1{\unskip}     \fi
\ifx \showLCCN     \undefined \def \showLCCN      #1{\unskip}     \fi
\ifx \shownote     \undefined \def \shownote      #1{#1}          \fi
\ifx \showarticletitle \undefined \def \showarticletitle #1{#1}   \fi
\ifx \showURL      \undefined \def \showURL       {\relax}        \fi
\providecommand\bibfield[2]{#2}
\providecommand\bibinfo[2]{#2}
\providecommand\natexlab[1]{#1}
\providecommand\showeprint[2][]{arXiv:#2}

\bibitem[\protect\citeauthoryear{Allamanis}{Allamanis}{2019}]%
        {allamanis2019adverse}
\bibfield{author}{\bibinfo{person}{Miltiadis Allamanis}.}
  \bibinfo{year}{2019}\natexlab{}.
\newblock \showarticletitle{The adverse effects of code duplication in machine
  learning models of code}. In \bibinfo{booktitle}{\emph{Proceedings of the
  2019 ACM SIGPLAN International Symposium on New Ideas, New Paradigms, and
  Reflections on Programming and Software}}. \bibinfo{pages}{143--153}.
\newblock


\bibitem[\protect\citeauthoryear{Alon, Zilberstein, Levy, and Yahav}{Alon
  et~al\mbox{.}}{2019}]%
        {alon2019code2vec}
\bibfield{author}{\bibinfo{person}{Uri Alon}, \bibinfo{person}{Meital
  Zilberstein}, \bibinfo{person}{Omer Levy}, {and} \bibinfo{person}{Eran
  Yahav}.} \bibinfo{year}{2019}\natexlab{}.
\newblock \showarticletitle{code2vec: learning distributed representations of
  code}.
\newblock \bibinfo{journal}{\emph{Proceedings of the ACM on Programming
  Languages}} \bibinfo{volume}{3}, \bibinfo{number}{{POPL}}
  (\bibinfo{year}{2019}), \bibinfo{pages}{40:1--40:29}.
\newblock
\urldef\tempurl%
\url{https://doi.org/10.1145/3290353}
\showDOI{\tempurl}


\bibitem[\protect\citeauthoryear{Bafna, Pramod, and Vaidya}{Bafna
  et~al\mbox{.}}{2016}]%
        {bafna2016document}
\bibfield{author}{\bibinfo{person}{Prafulla Bafna}, \bibinfo{person}{Dhanya
  Pramod}, {and} \bibinfo{person}{Anagha Vaidya}.}
  \bibinfo{year}{2016}\natexlab{}.
\newblock \showarticletitle{Document clustering: TF-IDF approach}. In
  \bibinfo{booktitle}{\emph{2016 International Conference on Electrical,
  Electronics, and Optimization Techniques (ICEEOT)}}. IEEE,
  \bibinfo{pages}{61--66}.
\newblock


\bibitem[\protect\citeauthoryear{Bettenburg, Just, Schr{\"o}ter, Weiss,
  Premraj, and Zimmermann}{Bettenburg et~al\mbox{.}}{2008}]%
        {bettenburg2008makes}
\bibfield{author}{\bibinfo{person}{Nicolas Bettenburg}, \bibinfo{person}{Sascha
  Just}, \bibinfo{person}{Adrian Schr{\"o}ter}, \bibinfo{person}{Cathrin
  Weiss}, \bibinfo{person}{Rahul Premraj}, {and} \bibinfo{person}{Thomas
  Zimmermann}.} \bibinfo{year}{2008}\natexlab{}.
\newblock \showarticletitle{What makes a good bug report?}. In
  \bibinfo{booktitle}{\emph{Proceedings of the 16th ACM SIGSOFT International
  Symposium on Foundations of software engineering}}.
  \bibinfo{pages}{308--318}.
\newblock


\bibitem[\protect\citeauthoryear{Chen, Kommrusch, Tufano, Pouchet, Poshyvanyk,
  and Monperrus}{Chen et~al\mbox{.}}{2019}]%
        {chen2019sequencer}
\bibfield{author}{\bibinfo{person}{Zimin Chen}, \bibinfo{person}{Steve
  Kommrusch}, \bibinfo{person}{Michele Tufano}, \bibinfo{person}{Louis-No{\"e}l
  Pouchet}, \bibinfo{person}{Denys Poshyvanyk}, {and} \bibinfo{person}{Martin
  Monperrus}.} \bibinfo{year}{2019}\natexlab{}.
\newblock \showarticletitle{Sequencer: Sequence-to-sequence learning for
  end-to-end program repair}.
\newblock \bibinfo{journal}{\emph{IEEE Transactions on Software Engineering}}
  \bibinfo{volume}{47}, \bibinfo{number}{9} (\bibinfo{year}{2019}),
  \bibinfo{pages}{1943--1959}.
\newblock


\bibitem[\protect\citeauthoryear{Church}{Church}{2017}]%
        {church2017word2vec}
\bibfield{author}{\bibinfo{person}{Kenneth~Ward Church}.}
  \bibinfo{year}{2017}\natexlab{}.
\newblock \showarticletitle{Word2Vec}.
\newblock \bibinfo{journal}{\emph{Natural Language Engineering}}
  \bibinfo{volume}{23}, \bibinfo{number}{1} (\bibinfo{year}{2017}),
  \bibinfo{pages}{155--162}.
\newblock


\bibitem[\protect\citeauthoryear{Devlin, Chang, Lee, and Toutanova}{Devlin
  et~al\mbox{.}}{2019}]%
        {devlin2019bert}
\bibfield{author}{\bibinfo{person}{Jacob Devlin}, \bibinfo{person}{Ming{-}Wei
  Chang}, \bibinfo{person}{Kenton Lee}, {and} \bibinfo{person}{Kristina
  Toutanova}.} \bibinfo{year}{2019}\natexlab{}.
\newblock \showarticletitle{{BERT:} Pre-training of Deep Bidirectional
  Transformers for Language Understanding}. In
  \bibinfo{booktitle}{\emph{Proceedings of the 2019 Conference of the North
  American Chapter of the Association for Computational Linguistics: Human
  Language Technologies}}. \bibinfo{pages}{4171--4186}.
\newblock
\urldef\tempurl%
\url{https://doi.org/10.18653/v1/n19-1423}
\showDOI{\tempurl}


\bibitem[\protect\citeauthoryear{Durieux, Madeiral, Martinez, and
  Abreu}{Durieux et~al\mbox{.}}{2019}]%
        {durieux2019empirical}
\bibfield{author}{\bibinfo{person}{Thomas Durieux}, \bibinfo{person}{Fernanda
  Madeiral}, \bibinfo{person}{Matias Martinez}, {and} \bibinfo{person}{Rui
  Abreu}.} \bibinfo{year}{2019}\natexlab{}.
\newblock \showarticletitle{Empirical Review of Java Program Repair Tools: A
  Large-Scale Experiment on 2,141 Bugs and 23,551 Repair Attempts}. In
  \bibinfo{booktitle}{\emph{Proceedings of the 27th ACM Joint Meeting on
  European Software Engineering Conference and Symposium on the Foundations of
  Software Engineering}}. \bibinfo{publisher}{ACM}, \bibinfo{pages}{302--313}.
\newblock
\urldef\tempurl%
\url{https://doi.org/10.1145/3338906.3338911}
\showDOI{\tempurl}


\bibitem[\protect\citeauthoryear{Elnaggar, Ding, Jones, Gibbs, Feher, Angerer,
  Severini, Matthes, and Rost}{Elnaggar et~al\mbox{.}}{2021}]%
        {elnaggar2021codetrans}
\bibfield{author}{\bibinfo{person}{Ahmed Elnaggar}, \bibinfo{person}{Wei Ding},
  \bibinfo{person}{Llion Jones}, \bibinfo{person}{Tom Gibbs},
  \bibinfo{person}{Tamas Feher}, \bibinfo{person}{Christoph Angerer},
  \bibinfo{person}{Silvia Severini}, \bibinfo{person}{Florian Matthes}, {and}
  \bibinfo{person}{Burkhard Rost}.} \bibinfo{year}{2021}\natexlab{}.
\newblock \showarticletitle{CodeTrans: Towards Cracking the Language of
  Silicon's Code Through Self-Supervised Deep Learning and High Performance
  Computing}.
\newblock \bibinfo{journal}{\emph{arXiv preprint arXiv:2104.02443}}
  (\bibinfo{year}{2021}).
\newblock


\bibitem[\protect\citeauthoryear{Fazzini, Prammer, d'Amorim, and Orso}{Fazzini
  et~al\mbox{.}}{2018}]%
        {fazzini2018automatically}
\bibfield{author}{\bibinfo{person}{Mattia Fazzini}, \bibinfo{person}{Martin
  Prammer}, \bibinfo{person}{Marcelo d'Amorim}, {and}
  \bibinfo{person}{Alessandro Orso}.} \bibinfo{year}{2018}\natexlab{}.
\newblock \showarticletitle{Automatically translating bug reports into test
  cases for mobile apps}. In \bibinfo{booktitle}{\emph{Proceedings of the 27th
  ACM SIGSOFT International Symposium on Software Testing and Analysis}}.
  \bibinfo{pages}{141--152}.
\newblock


\bibitem[\protect\citeauthoryear{Fraser and Arcuri}{Fraser and Arcuri}{2011}]%
        {fraser2011evoSuite}
\bibfield{author}{\bibinfo{person}{Gordon Fraser} {and} \bibinfo{person}{Andrea
  Arcuri}.} \bibinfo{year}{2011}\natexlab{}.
\newblock \showarticletitle{EvoSuite: Automatic test suite generation for
  object-oriented software}. In \bibinfo{booktitle}{\emph{SIGSOFT/FSE'11 19th
  ACM SIGSOFT Symposium on the Foundations of Software Engineering (FSE-19) and
  ESEC'11: 13rd European Software Engineering Conference (ESEC-13), Szeged,
  Hungary, September 5-9, 2011}}.
\newblock


\bibitem[\protect\citeauthoryear{Gao, Xia, Lo, Grundy, and Li}{Gao
  et~al\mbox{.}}{2021}]%
        {gao2021code2que}
\bibfield{author}{\bibinfo{person}{Zhipeng Gao}, \bibinfo{person}{Xin Xia},
  \bibinfo{person}{David Lo}, \bibinfo{person}{John Grundy}, {and}
  \bibinfo{person}{Yuan-Fang Li}.} \bibinfo{year}{2021}\natexlab{}.
\newblock \showarticletitle{Code2Que: A tool for improving question titles from
  mined code snippets in stack overflow}. In
  \bibinfo{booktitle}{\emph{Proceedings of the 29th ACM Joint Meeting on
  European Software Engineering Conference and Symposium on the Foundations of
  Software Engineering}}. \bibinfo{pages}{1525--1529}.
\newblock


\bibitem[\protect\citeauthoryear{Gazzola, Micucci, and Mariani}{Gazzola
  et~al\mbox{.}}{2017}]%
        {gazzola2017automatic}
\bibfield{author}{\bibinfo{person}{Luca Gazzola}, \bibinfo{person}{Daniela
  Micucci}, {and} \bibinfo{person}{Leonardo Mariani}.}
  \bibinfo{year}{2017}\natexlab{}.
\newblock \showarticletitle{Automatic software repair: A survey}.
\newblock \bibinfo{journal}{\emph{IEEE Transactions on Software Engineering}}
  \bibinfo{volume}{45}, \bibinfo{number}{1} (\bibinfo{year}{2017}),
  \bibinfo{pages}{34--67}.
\newblock


\bibitem[\protect\citeauthoryear{Hindle, Barr, Su, Gabel, and Devanbu}{Hindle
  et~al\mbox{.}}{2012}]%
        {hindle2012naturalness}
\bibfield{author}{\bibinfo{person}{Abram Hindle}, \bibinfo{person}{Earl~T.
  Barr}, \bibinfo{person}{Zhendong Su}, \bibinfo{person}{Mark Gabel}, {and}
  \bibinfo{person}{Premkumar~T. Devanbu}.} \bibinfo{year}{2012}\natexlab{}.
\newblock \showarticletitle{On the naturalness of software}. In
  \bibinfo{booktitle}{\emph{Proceedings of the 34th International Conference on
  Software Engineering}}. IEEE, \bibinfo{pages}{837--847}.
\newblock
\urldef\tempurl%
\url{https://doi.org/10.1109/ICSE.2012.6227135}
\showDOI{\tempurl}


\bibitem[\protect\citeauthoryear{Hoang, Kang, Lawall, and Lo}{Hoang
  et~al\mbox{.}}{2020}]%
        {hoang2020cc2vec}
\bibfield{author}{\bibinfo{person}{Thong Hoang}, \bibinfo{person}{Hong~Jin
  Kang}, \bibinfo{person}{Julia Lawall}, {and} \bibinfo{person}{David Lo}.}
  \bibinfo{year}{2020}\natexlab{}.
\newblock \showarticletitle{{CC2Vec:} Distributed Representations of Code
  Changes}. In \bibinfo{booktitle}{\emph{Proceedings of the 42nd International
  Conference on Software Engineering}}. ACM, \bibinfo{pages}{518--529}.
\newblock
\urldef\tempurl%
\url{https://doi.org/10.1145/3377811.3380361}
\showDOI{\tempurl}


\bibitem[\protect\citeauthoryear{Hossin and Sulaiman}{Hossin and
  Sulaiman}{2015}]%
        {hossin2015review}
\bibfield{author}{\bibinfo{person}{Mohammad Hossin} {and}
  \bibinfo{person}{Md~Nasir Sulaiman}.} \bibinfo{year}{2015}\natexlab{}.
\newblock \showarticletitle{A review on evaluation metrics for data
  classification evaluations}.
\newblock \bibinfo{journal}{\emph{International journal of data mining \&
  knowledge management process}} \bibinfo{volume}{5}, \bibinfo{number}{2}
  (\bibinfo{year}{2015}), \bibinfo{pages}{1}.
\newblock


\bibitem[\protect\citeauthoryear{Irolla and Dey}{Irolla and Dey}{2018}]%
        {irolla2018duplication}
\bibfield{author}{\bibinfo{person}{Paul Irolla} {and}
  \bibinfo{person}{Alexandre Dey}.} \bibinfo{year}{2018}\natexlab{}.
\newblock \showarticletitle{The duplication issue within the drebin dataset}.
\newblock \bibinfo{journal}{\emph{Journal of Computer Virology and Hacking
  Techniques}} \bibinfo{volume}{14}, \bibinfo{number}{3}
  (\bibinfo{year}{2018}), \bibinfo{pages}{245--249}.
\newblock


\bibitem[\protect\citeauthoryear{Jeni, Cohn, and De~La~Torre}{Jeni
  et~al\mbox{.}}{2013}]%
        {jeni2013facing}
\bibfield{author}{\bibinfo{person}{L{\'a}szl{\'o}~A Jeni},
  \bibinfo{person}{Jeffrey~F Cohn}, {and} \bibinfo{person}{Fernando
  De~La~Torre}.} \bibinfo{year}{2013}\natexlab{}.
\newblock \showarticletitle{Facing imbalanced data--recommendations for the use
  of performance metrics}. In \bibinfo{booktitle}{\emph{2013 Humaine
  association conference on affective computing and intelligent interaction}}.
  IEEE, \bibinfo{pages}{245--251}.
\newblock


\bibitem[\protect\citeauthoryear{Jiang, Lutellier, and Tan}{Jiang
  et~al\mbox{.}}{2021}]%
        {jiang2021cure}
\bibfield{author}{\bibinfo{person}{Nan Jiang}, \bibinfo{person}{Thibaud
  Lutellier}, {and} \bibinfo{person}{Lin Tan}.}
  \bibinfo{year}{2021}\natexlab{}.
\newblock \showarticletitle{CURE: Code-aware neural machine translation for
  automatic program repair}. In \bibinfo{booktitle}{\emph{2021 IEEE/ACM 43rd
  International Conference on Software Engineering (ICSE)}}. IEEE,
  \bibinfo{pages}{1161--1173}.
\newblock


\bibitem[\protect\citeauthoryear{Joulin, Grave, Bojanowski, Douze, J{\'e}gou,
  and Mikolov}{Joulin et~al\mbox{.}}{2016}]%
        {joulin2016fasttext}
\bibfield{author}{\bibinfo{person}{Armand Joulin}, \bibinfo{person}{Edouard
  Grave}, \bibinfo{person}{Piotr Bojanowski}, \bibinfo{person}{Matthijs Douze},
  \bibinfo{person}{H{\'e}rve J{\'e}gou}, {and} \bibinfo{person}{Tomas
  Mikolov}.} \bibinfo{year}{2016}\natexlab{}.
\newblock \showarticletitle{FastText.zip: Compressing text classification
  models}.
\newblock \bibinfo{journal}{\emph{arXiv preprint arXiv:1612.03651}}
  (\bibinfo{year}{2016}).
\newblock


\bibitem[\protect\citeauthoryear{Just, Jalali, and Ernst}{Just
  et~al\mbox{.}}{2014}]%
        {just2014defects4j}
\bibfield{author}{\bibinfo{person}{Ren{\'e} Just}, \bibinfo{person}{Darioush
  Jalali}, {and} \bibinfo{person}{Michael~D Ernst}.}
  \bibinfo{year}{2014}\natexlab{}.
\newblock \showarticletitle{{Defects4J}: A database of existing faults to
  enable controlled testing studies for Java programs}. In
  \bibinfo{booktitle}{\emph{Proceedings of the 23rd International Symposium on
  Software Testing and Analysis}}. ACM, \bibinfo{pages}{437--440}.
\newblock
\urldef\tempurl%
\url{https://doi.org/10.1145/2610384.2628055}
\showDOI{\tempurl}


\bibitem[\protect\citeauthoryear{Khanfir, Koyuncu, Papadakis, Cordy,
  Bissyand{\'e}, Klein, and Traon}{Khanfir et~al\mbox{.}}{2020}]%
        {khanfir2020ibir}
\bibfield{author}{\bibinfo{person}{Ahmed Khanfir}, \bibinfo{person}{Anil
  Koyuncu}, \bibinfo{person}{Mike Papadakis}, \bibinfo{person}{Maxime Cordy},
  \bibinfo{person}{Tegawend{\'e}~F Bissyand{\'e}}, \bibinfo{person}{Jacques
  Klein}, {and} \bibinfo{person}{Yves~Le Traon}.}
  \bibinfo{year}{2020}\natexlab{}.
\newblock \showarticletitle{Ibir: Bug report driven fault injection}.
\newblock \bibinfo{journal}{\emph{arXiv preprint arXiv:2012.06506}}
  (\bibinfo{year}{2020}).
\newblock


\bibitem[\protect\citeauthoryear{Kim, Kim, and Lee}{Kim et~al\mbox{.}}{2021}]%
        {kim2021denchmark}
\bibfield{author}{\bibinfo{person}{Misoo Kim}, \bibinfo{person}{Youngkyoung
  Kim}, {and} \bibinfo{person}{Eunseok Lee}.} \bibinfo{year}{2021}\natexlab{}.
\newblock \showarticletitle{Denchmark: A Bug Benchmark of Deep Learning-related
  Software}. In \bibinfo{booktitle}{\emph{2021 IEEE/ACM 18th International
  Conference on Mining Software Repositories (MSR)}}. IEEE,
  \bibinfo{pages}{540--544}.
\newblock


\bibitem[\protect\citeauthoryear{Koyuncu, Liu, Bissyand{\'e}, Kim, Monperrus,
  Klein, and Le~Traon}{Koyuncu et~al\mbox{.}}{2019}]%
        {koyuncu2019ifixr}
\bibfield{author}{\bibinfo{person}{Anil Koyuncu}, \bibinfo{person}{Kui Liu},
  \bibinfo{person}{Tegawend{\'e}~F. Bissyand{\'e}}, \bibinfo{person}{Dongsun
  Kim}, \bibinfo{person}{Martin Monperrus}, \bibinfo{person}{Jacques Klein},
  {and} \bibinfo{person}{Yves Le~Traon}.} \bibinfo{year}{2019}\natexlab{}.
\newblock \showarticletitle{{iFixR}: Bug Report driven Program Repair}. In
  \bibinfo{booktitle}{\emph{Proceedings of the 27the ACM Joint European
  Software Engineering Conference and Symposium on the Foundations of Software
  Engineering}}. ACM, \bibinfo{pages}{314--325}.
\newblock
\urldef\tempurl%
\url{https://doi.org/10.1145/3338906.3338935}
\showDOI{\tempurl}


\bibitem[\protect\citeauthoryear{Li, Wang, and Nguyen}{Li
  et~al\mbox{.}}{2020}]%
        {li2020dlfix}
\bibfield{author}{\bibinfo{person}{Yi Li}, \bibinfo{person}{Shaohua Wang},
  {and} \bibinfo{person}{Tien~N Nguyen}.} \bibinfo{year}{2020}\natexlab{}.
\newblock \showarticletitle{Dlfix: Context-based code transformation learning
  for automated program repair}. In \bibinfo{booktitle}{\emph{Proceedings of
  the ACM/IEEE 42nd International Conference on Software Engineering}}.
  \bibinfo{pages}{602--614}.
\newblock


\bibitem[\protect\citeauthoryear{Lin, Wang, Wen, and Mao}{Lin
  et~al\mbox{.}}{2022}]%
        {lin2022context}
\bibfield{author}{\bibinfo{person}{Bo Lin}, \bibinfo{person}{Shangwen Wang},
  \bibinfo{person}{Ming Wen}, {and} \bibinfo{person}{Xiaoguang Mao}.}
  \bibinfo{year}{2022}\natexlab{}.
\newblock \showarticletitle{Context-Aware Code Change Embedding for Better
  Patch Correctness Assessment}.
\newblock \bibinfo{journal}{\emph{ACM Trans. Softw. Eng. Methodol.}}
  \bibinfo{volume}{31}, \bibinfo{number}{3}, Article \bibinfo{articleno}{51}
  (\bibinfo{date}{may} \bibinfo{year}{2022}), \bibinfo{numpages}{29}~pages.
\newblock
\showISSN{1049-331X}
\urldef\tempurl%
\url{https://doi.org/10.1145/3505247}
\showDOI{\tempurl}


\bibitem[\protect\citeauthoryear{Liu, Yang, Tan, and Hafiz}{Liu
  et~al\mbox{.}}{2013}]%
        {liu2013r2fix}
\bibfield{author}{\bibinfo{person}{Chen Liu}, \bibinfo{person}{Jinqiu Yang},
  \bibinfo{person}{Lin Tan}, {and} \bibinfo{person}{Munawar Hafiz}.}
  \bibinfo{year}{2013}\natexlab{}.
\newblock \showarticletitle{R2Fix: Automatically generating bug fixes from bug
  reports}. In \bibinfo{booktitle}{\emph{2013 IEEE Sixth international
  conference on software testing, verification and validation}}. IEEE,
  \bibinfo{pages}{282--291}.
\newblock


\bibitem[\protect\citeauthoryear{Liu, Koyuncu, Kim, and Bissyand{\'e}}{Liu
  et~al\mbox{.}}{2019}]%
        {liu2019avatar}
\bibfield{author}{\bibinfo{person}{Kui Liu}, \bibinfo{person}{Anil Koyuncu},
  \bibinfo{person}{Dongsun Kim}, {and} \bibinfo{person}{Tegawend{\'e}~F
  Bissyand{\'e}}.} \bibinfo{year}{2019}\natexlab{}.
\newblock \showarticletitle{{AVATAR:} Fixing semantic bugs with fix patterns of
  static analysis violations}. In \bibinfo{booktitle}{\emph{Proceedings of the
  26th IEEE International Conference on Software Analysis, Evolution and
  Reengineering}}. IEEE, \bibinfo{pages}{456--467}.
\newblock
\urldef\tempurl%
\url{https://doi.org/10.1109/SANER.2019.8667970}
\showDOI{\tempurl}


\bibitem[\protect\citeauthoryear{Liu, Li, Koyuncu, Kim, Liu, Klein, and
  Bissyand{\'e}}{Liu et~al\mbox{.}}{2021}]%
        {liu2021critical}
\bibfield{author}{\bibinfo{person}{Kui Liu}, \bibinfo{person}{Li Li},
  \bibinfo{person}{Anil Koyuncu}, \bibinfo{person}{Dongsun Kim},
  \bibinfo{person}{Zhe Liu}, \bibinfo{person}{Jacques Klein}, {and}
  \bibinfo{person}{Tegawend{\'e}~F. Bissyand{\'e}}.}
  \bibinfo{year}{2021}\natexlab{}.
\newblock \showarticletitle{A Critical Review on the Evaluation of Automated
  Program Repair Systems}.
\newblock \bibinfo{journal}{\emph{Journal of Systems and Software}}
  \bibinfo{volume}{171} (\bibinfo{year}{2021}), \bibinfo{pages}{110817}.
\newblock
\urldef\tempurl%
\url{https://doi.org/10.1016/j.jss.2020.110817}
\showDOI{\tempurl}


\bibitem[\protect\citeauthoryear{Liu, Wang, Koyuncu, Kim, Bissyandé, Kim, Wu,
  Klein, Mao, and Traon}{Liu et~al\mbox{.}}{2020}]%
        {liu2020efficiency}
\bibfield{author}{\bibinfo{person}{Kui Liu}, \bibinfo{person}{Shangwen Wang},
  \bibinfo{person}{Anil Koyuncu}, \bibinfo{person}{Kisub Kim},
  \bibinfo{person}{Tegawendé~F. Bissyandé}, \bibinfo{person}{Dongsun Kim},
  \bibinfo{person}{Peng Wu}, \bibinfo{person}{Jacques Klein},
  \bibinfo{person}{Xiaoguang Mao}, {and} \bibinfo{person}{Yves~Le Traon}.}
  \bibinfo{year}{2020}\natexlab{}.
\newblock \showarticletitle{On the Efficiency of Test Suite based Program
  Repair: A Systematic Assessment of 16 Automated Repair Systems for Java
  Programs}. In \bibinfo{booktitle}{\emph{Proceedings of the 42nd International
  Conference on Software Engineering}}. ACM, \bibinfo{pages}{615--627}.
\newblock
\urldef\tempurl%
\url{https://doi.org/10.1145/3377811.3380338}
\showDOI{\tempurl}


\bibitem[\protect\citeauthoryear{Long and Rinard}{Long and Rinard}{2016}]%
        {long2016analysis}
\bibfield{author}{\bibinfo{person}{Fan Long} {and} \bibinfo{person}{Martin
  Rinard}.} \bibinfo{year}{2016}\natexlab{}.
\newblock \showarticletitle{An analysis of the search spaces for generate and
  validate patch generation systems}. In \bibinfo{booktitle}{\emph{Proceedings
  of the 38th International Conference on Software Engineering}}. IEEE,
  \bibinfo{pages}{702--713}.
\newblock
\urldef\tempurl%
\url{https://doi.org/10.1145/2884781.2884872}
\showDOI{\tempurl}


\bibitem[\protect\citeauthoryear{Lutellier, Pham, Pang, Li, Wei, and
  Tan}{Lutellier et~al\mbox{.}}{2020}]%
        {lutellier2020coconut}
\bibfield{author}{\bibinfo{person}{Thibaud Lutellier},
  \bibinfo{person}{Hung~Viet Pham}, \bibinfo{person}{Lawrence Pang},
  \bibinfo{person}{Yitong Li}, \bibinfo{person}{Moshi Wei}, {and}
  \bibinfo{person}{Lin Tan}.} \bibinfo{year}{2020}\natexlab{}.
\newblock \showarticletitle{Coconut: combining context-aware neural translation
  models using ensemble for program repair}. In
  \bibinfo{booktitle}{\emph{Proceedings of the 29th ACM SIGSOFT international
  symposium on software testing and analysis}}. \bibinfo{pages}{101--114}.
\newblock


\bibitem[\protect\citeauthoryear{Madeiral, Urli, Maia, and Monperrus}{Madeiral
  et~al\mbox{.}}{2019}]%
        {madeiral2019bears}
\bibfield{author}{\bibinfo{person}{Fernanda Madeiral}, \bibinfo{person}{Simon
  Urli}, \bibinfo{person}{Marcelo Maia}, {and} \bibinfo{person}{Martin
  Monperrus}.} \bibinfo{year}{2019}\natexlab{}.
\newblock \showarticletitle{{BEARS:} An Extensible Java Bug Benchmark for
  Automatic Program Repair Studies}. In \bibinfo{booktitle}{\emph{Proceedings
  of the 26th International Conference on Software Analysis, Evolution and
  Reengineering}}. IEEE, \bibinfo{pages}{468--478}.
\newblock
\urldef\tempurl%
\url{https://doi.org/10.1109/SANER.2019.8667991}
\showDOI{\tempurl}


\bibitem[\protect\citeauthoryear{Mashhadi and Hemmati}{Mashhadi and
  Hemmati}{2021}]%
        {mashhadi2021applying}
\bibfield{author}{\bibinfo{person}{Ehsan Mashhadi} {and} \bibinfo{person}{Hadi
  Hemmati}.} \bibinfo{year}{2021}\natexlab{}.
\newblock \showarticletitle{Applying codebert for automated program repair of
  java simple bugs}. In \bibinfo{booktitle}{\emph{2021 IEEE/ACM 18th
  International Conference on Mining Software Repositories (MSR)}}. IEEE,
  \bibinfo{pages}{505--509}.
\newblock


\bibitem[\protect\citeauthoryear{McKnight and Najab}{McKnight and
  Najab}{2010}]%
        {mcknight2010mann}
\bibfield{author}{\bibinfo{person}{Patrick~E McKnight} {and}
  \bibinfo{person}{Julius Najab}.} \bibinfo{year}{2010}\natexlab{}.
\newblock \showarticletitle{Mann-Whitney U Test}.
\newblock \bibinfo{journal}{\emph{The Corsini encyclopedia of psychology}}
  (\bibinfo{year}{2010}), \bibinfo{pages}{1--1}.
\newblock


\bibitem[\protect\citeauthoryear{Mikolov, Chen, Corrado, and Dean}{Mikolov
  et~al\mbox{.}}{2013}]%
        {mikolov2013efficient}
\bibfield{author}{\bibinfo{person}{Tomas Mikolov}, \bibinfo{person}{Kai Chen},
  \bibinfo{person}{Greg Corrado}, {and} \bibinfo{person}{Jeffrey Dean}.}
  \bibinfo{year}{2013}\natexlab{}.
\newblock \showarticletitle{Efficient estimation of word representations in
  vector space}.
\newblock \bibinfo{journal}{\emph{arXiv preprint arXiv:1301.3781}}
  (\bibinfo{year}{2013}).
\newblock


\bibitem[\protect\citeauthoryear{Min, Seo, and Hajishirzi}{Min
  et~al\mbox{.}}{2017}]%
        {min2017question}
\bibfield{author}{\bibinfo{person}{Sewon Min}, \bibinfo{person}{Minjoon Seo},
  {and} \bibinfo{person}{Hannaneh Hajishirzi}.}
  \bibinfo{year}{2017}\natexlab{}.
\newblock \showarticletitle{Question answering through transfer learning from
  large fine-grained supervision data}.
\newblock \bibinfo{journal}{\emph{arXiv preprint arXiv:1702.02171}}
  (\bibinfo{year}{2017}).
\newblock


\bibitem[\protect\citeauthoryear{Monperrus}{Monperrus}{2020}]%
        {monperrus2020living}
\bibfield{author}{\bibinfo{person}{Martin Monperrus}.}
  \bibinfo{year}{2020}\natexlab{}.
\newblock \showarticletitle{The living review on automated program repair}.
\newblock  (\bibinfo{year}{2020}).
\newblock


\bibitem[\protect\citeauthoryear{Pacheco and Ernst}{Pacheco and Ernst}{2007}]%
        {pacheco2007randoop}
\bibfield{author}{\bibinfo{person}{Carlos Pacheco} {and}
  \bibinfo{person}{Michael~D. Ernst}.} \bibinfo{year}{2007}\natexlab{}.
\newblock \showarticletitle{Randoop: feedback-directed random testing for
  Java}. In \bibinfo{booktitle}{\emph{In OOPSLA ’07 Companion}}.
  \bibinfo{publisher}{ACM}, \bibinfo{pages}{815--816}.
\newblock


\bibitem[\protect\citeauthoryear{Pennington, Socher, and Manning}{Pennington
  et~al\mbox{.}}{2014}]%
        {pennington2014glove}
\bibfield{author}{\bibinfo{person}{Jeffrey Pennington},
  \bibinfo{person}{Richard Socher}, {and} \bibinfo{person}{Christopher~D
  Manning}.} \bibinfo{year}{2014}\natexlab{}.
\newblock \showarticletitle{Glove: Global vectors for word representation}. In
  \bibinfo{booktitle}{\emph{Proceedings of the 2014 conference on empirical
  methods in natural language processing (EMNLP)}}.
  \bibinfo{pages}{1532--1543}.
\newblock


\bibitem[\protect\citeauthoryear{Qi, Long, Achour, and Rinard}{Qi
  et~al\mbox{.}}{2015}]%
        {qi2015analysis}
\bibfield{author}{\bibinfo{person}{Zichao Qi}, \bibinfo{person}{Fan Long},
  \bibinfo{person}{Sara Achour}, {and} \bibinfo{person}{Martin Rinard}.}
  \bibinfo{year}{2015}\natexlab{}.
\newblock \showarticletitle{An analysis of patch plausibility and correctness
  for generate-and-validate patch generation systems}. In
  \bibinfo{booktitle}{\emph{Proceedings of the 24th International Symposium on
  Software Testing and Analysis}}. ACM, \bibinfo{pages}{24--36}.
\newblock
\urldef\tempurl%
\url{https://doi.org/10.1145/2771783.2771791}
\showDOI{\tempurl}


\bibitem[\protect\citeauthoryear{Rajpurkar, Zhang, Lopyrev, and
  Liang}{Rajpurkar et~al\mbox{.}}{2016}]%
        {rajpurkar2016squad}
\bibfield{author}{\bibinfo{person}{Pranav Rajpurkar}, \bibinfo{person}{Jian
  Zhang}, \bibinfo{person}{Konstantin Lopyrev}, {and} \bibinfo{person}{Percy
  Liang}.} \bibinfo{year}{2016}\natexlab{}.
\newblock \showarticletitle{Squad: 100,000+ questions for machine comprehension
  of text}.
\newblock \bibinfo{journal}{\emph{arXiv preprint arXiv:1606.05250}}
  (\bibinfo{year}{2016}).
\newblock


\bibitem[\protect\citeauthoryear{Saha, Lyu, Lam, Yoshida, and Prasad}{Saha
  et~al\mbox{.}}{2018}]%
        {saha2018bugs}
\bibfield{author}{\bibinfo{person}{Ripon Saha}, \bibinfo{person}{Yingjun Lyu},
  \bibinfo{person}{Wing Lam}, \bibinfo{person}{Hiroaki Yoshida}, {and}
  \bibinfo{person}{Mukul Prasad}.} \bibinfo{year}{2018}\natexlab{}.
\newblock \showarticletitle{Bugs.jar: A large-scale, diverse dataset of
  real-world java bugs}. In \bibinfo{booktitle}{\emph{Proceedings of the 15th
  IEEE/ACM International Conference on Mining Software Repositories}}. ACM,
  \bibinfo{pages}{10--13}.
\newblock
\urldef\tempurl%
\url{https://doi.org/10.1145/3196398.3196473}
\showDOI{\tempurl}


\bibitem[\protect\citeauthoryear{Shariffdeen, Noller, Grunske, and
  Roychoudhury}{Shariffdeen et~al\mbox{.}}{2021}]%
        {shariffdeen2021concolic}
\bibfield{author}{\bibinfo{person}{Ridwan Shariffdeen}, \bibinfo{person}{Yannic
  Noller}, \bibinfo{person}{Lars Grunske}, {and} \bibinfo{person}{Abhik
  Roychoudhury}.} \bibinfo{year}{2021}\natexlab{}.
\newblock \showarticletitle{Concolic program repair}. In
  \bibinfo{booktitle}{\emph{Proceedings of the 42nd ACM SIGPLAN International
  Conference on Programming Language Design and Implementation}}.
  \bibinfo{pages}{390--405}.
\newblock


\bibitem[\protect\citeauthoryear{Tan, Santos, Xiang, and Zhou}{Tan
  et~al\mbox{.}}{2015}]%
        {tan2015lstm}
\bibfield{author}{\bibinfo{person}{Ming Tan}, \bibinfo{person}{Cicero~dos
  Santos}, \bibinfo{person}{Bing Xiang}, {and} \bibinfo{person}{Bowen Zhou}.}
  \bibinfo{year}{2015}\natexlab{}.
\newblock \showarticletitle{Lstm-based deep learning models for non-factoid
  answer selection}.
\newblock \bibinfo{journal}{\emph{arXiv preprint arXiv:1511.04108}}
  (\bibinfo{year}{2015}).
\newblock


\bibitem[\protect\citeauthoryear{Tan, Yoshida, Prasad, and Roychoudhury}{Tan
  et~al\mbox{.}}{2016}]%
        {tan2016anti}
\bibfield{author}{\bibinfo{person}{Shin~Hwei Tan}, \bibinfo{person}{Hiroaki
  Yoshida}, \bibinfo{person}{Mukul~R. Prasad}, {and} \bibinfo{person}{Abhik
  Roychoudhury}.} \bibinfo{year}{2016}\natexlab{}.
\newblock \showarticletitle{Anti-patterns in search-based program repair}. In
  \bibinfo{booktitle}{\emph{Proceedings of the 2016 24th ACM SIGSOFT
  International Symposium on Foundations of Software Engineering}}.
  \bibinfo{pages}{727--738}.
\newblock


\bibitem[\protect\citeauthoryear{Tian, Li, Pian, Kabore, Liu, Habib, Klein, and
  Bissyand{\'e}}{Tian et~al\mbox{.}}{2022a}]%
        {tian2022predicting}
\bibfield{author}{\bibinfo{person}{Haoye Tian}, \bibinfo{person}{Yinghua Li},
  \bibinfo{person}{Weiguo Pian}, \bibinfo{person}{Abdoul~Kader Kabore},
  \bibinfo{person}{Kui Liu}, \bibinfo{person}{Andrew Habib},
  \bibinfo{person}{Jacques Klein}, {and} \bibinfo{person}{Tegawend{\'e}~F
  Bissyand{\'e}}.} \bibinfo{year}{2022}\natexlab{a}.
\newblock \showarticletitle{Predicting Patch Correctness Based on the
  Similarity of Failing Test Cases}.
\newblock \bibinfo{journal}{\emph{ACM Transactions on Software Engineering and
  Methodology}} (\bibinfo{year}{2022}).
\newblock
\urldef\tempurl%
\url{https://doi.org/10.1145/3511096}
\showDOI{\tempurl}


\bibitem[\protect\citeauthoryear{Tian, Liu, Kabor{\'e}, Koyuncu, Li, Klein, and
  Bissyand{\'e}}{Tian et~al\mbox{.}}{2020}]%
        {tian2020evaluating}
\bibfield{author}{\bibinfo{person}{Haoye Tian}, \bibinfo{person}{Kui Liu},
  \bibinfo{person}{Abdoul~Kader Kabor{\'e}}, \bibinfo{person}{Anil Koyuncu},
  \bibinfo{person}{Li Li}, \bibinfo{person}{Jacques Klein}, {and}
  \bibinfo{person}{Tegawend{\'e}~F Bissyand{\'e}}.}
  \bibinfo{year}{2020}\natexlab{}.
\newblock \showarticletitle{Evaluating representation learning of code changes
  for predicting patch correctness in program repair}. In
  \bibinfo{booktitle}{\emph{2020 35th IEEE/ACM International Conference on
  Automated Software Engineering (ASE)}}. IEEE, \bibinfo{pages}{981--992}.
\newblock
\urldef\tempurl%
\url{https://doi.org/10.1145/3324884.3416532}
\showDOI{\tempurl}


\bibitem[\protect\citeauthoryear{Tian, Liu, Li, Kabor{\'e}, Koyuncu, Habib, Li,
  Wen, Klein, and Bissyand{\'e}}{Tian et~al\mbox{.}}{2022b}]%
        {tian2022best}
\bibfield{author}{\bibinfo{person}{Haoye Tian}, \bibinfo{person}{Kui Liu},
  \bibinfo{person}{Yinghua Li}, \bibinfo{person}{Abdoul~Kader Kabor{\'e}},
  \bibinfo{person}{Anil Koyuncu}, \bibinfo{person}{Andrew Habib},
  \bibinfo{person}{Li Li}, \bibinfo{person}{Junhao Wen},
  \bibinfo{person}{Jacques Klein}, {and} \bibinfo{person}{Tegawend{\'e}~F
  Bissyand{\'e}}.} \bibinfo{year}{2022}\natexlab{b}.
\newblock \showarticletitle{The Best of Both Worlds: Combining Learned
  Embeddings with Engineered Features for Accurate Prediction of Correct
  Patches}.
\newblock \bibinfo{journal}{\emph{arXiv preprint arXiv:2203.08912}}
  (\bibinfo{year}{2022}).
\newblock


\bibitem[\protect\citeauthoryear{Tufano, Watson, Bavota, Di~Penta, White, and
  Poshyvanyk}{Tufano et~al\mbox{.}}{2019}]%
        {tufano2019empirical}
\bibfield{author}{\bibinfo{person}{Michele Tufano}, \bibinfo{person}{Cody
  Watson}, \bibinfo{person}{Gabriele Bavota}, \bibinfo{person}{Massimiliano
  Di~Penta}, \bibinfo{person}{Martin White}, {and} \bibinfo{person}{Denys
  Poshyvanyk}.} \bibinfo{year}{2019}\natexlab{}.
\newblock \showarticletitle{An empirical study on learning bug-fixing patches
  in the wild via neural machine translation}.
\newblock \bibinfo{journal}{\emph{ACM Transactions on Software Engineering and
  Methodology}} \bibinfo{volume}{28}, \bibinfo{number}{4}
  (\bibinfo{year}{2019}), \bibinfo{pages}{19:1--19:29}.
\newblock
\urldef\tempurl%
\url{https://doi.org/10.1145/3340544}
\showDOI{\tempurl}


\bibitem[\protect\citeauthoryear{Wang, Wen, Chen, Yi, and Mao}{Wang
  et~al\mbox{.}}{2019}]%
        {wang2019different}
\bibfield{author}{\bibinfo{person}{Shangwen Wang}, \bibinfo{person}{Ming Wen},
  \bibinfo{person}{Liqian Chen}, \bibinfo{person}{Xin Yi}, {and}
  \bibinfo{person}{Xiaoguang Mao}.} \bibinfo{year}{2019}\natexlab{}.
\newblock \showarticletitle{How Different Is It Between Machine-Generated and
  Developer-Provided Patches? An Empirical Study on The Correct Patches
  Generated by Automated Program Repair Techniques}. In
  \bibinfo{booktitle}{\emph{Proceedings of the 13th International Symposium on
  Empirical Software Engineering and Measurement}}.
  \bibinfo{publisher}{{IEEE}}, \bibinfo{pages}{1--12}.
\newblock
\urldef\tempurl%
\url{https://doi.org/10.1109/ESEM.2019.8870172}
\showDOI{\tempurl}


\bibitem[\protect\citeauthoryear{Wang, Wen, Lin, Wu, Qin, Zou, Mao, and
  Jin}{Wang et~al\mbox{.}}{2020}]%
        {wang2020automated}
\bibfield{author}{\bibinfo{person}{Shangwen Wang}, \bibinfo{person}{Ming Wen},
  \bibinfo{person}{Bo Lin}, \bibinfo{person}{Hongjun Wu},
  \bibinfo{person}{Yihao Qin}, \bibinfo{person}{Deqing Zou},
  \bibinfo{person}{Xiaoguang Mao}, {and} \bibinfo{person}{Hai Jin}.}
  \bibinfo{year}{2020}\natexlab{}.
\newblock \showarticletitle{Automated Patch Correctness Assessment: How Far are
  We?}. In \bibinfo{booktitle}{\emph{Proceedings of the 35th IEEE/ACM
  International Conference on Automated Software Engineering (ASE)}}. ACM,
  \bibinfo{pages}{968--980}.
\newblock
\urldef\tempurl%
\url{https://doi.org/10.1145/3324884.3416590}
\showDOI{\tempurl}


\bibitem[\protect\citeauthoryear{Xin and Reiss}{Xin and Reiss}{2017}]%
        {xin2017identifying}
\bibfield{author}{\bibinfo{person}{Qi Xin} {and} \bibinfo{person}{Steven~P
  Reiss}.} \bibinfo{year}{2017}\natexlab{}.
\newblock \showarticletitle{Identifying test-suite-overfitted patches through
  test case generation}. In \bibinfo{booktitle}{\emph{Proceedings of the 26th
  ACM SIGSOFT International Symposium on Software Testing and Analysis}}. ACM,
  \bibinfo{pages}{226--236}.
\newblock
\urldef\tempurl%
\url{https://doi.org/10.1145/3092703.3092718}
\showDOI{\tempurl}


\bibitem[\protect\citeauthoryear{Xiong, Liu, Zeng, Zhang, and Huang}{Xiong
  et~al\mbox{.}}{2018}]%
        {xiong2018identifying}
\bibfield{author}{\bibinfo{person}{Yingfei Xiong}, \bibinfo{person}{Xinyuan
  Liu}, \bibinfo{person}{Muhan Zeng}, \bibinfo{person}{Lu Zhang}, {and}
  \bibinfo{person}{Gang Huang}.} \bibinfo{year}{2018}\natexlab{}.
\newblock \showarticletitle{Identifying patch correctness in test-based program
  repair}. In \bibinfo{booktitle}{\emph{Proceedings of the 40th International
  Conference on Software Engineering}}. ACM, \bibinfo{pages}{789--799}.
\newblock
\urldef\tempurl%
\url{https://doi.org/10.1145/3180155.3180182}
\showDOI{\tempurl}


\bibitem[\protect\citeauthoryear{Yan, Liu, Niu, Li, Zhe, Liu, Klein, and
  F.~Bissyandé}{Yan et~al\mbox{.}}{2022}]%
        {yan2022crex}
\bibfield{author}{\bibinfo{person}{Dapeng Yan}, \bibinfo{person}{Kui Liu},
  \bibinfo{person}{Yuqing Niu}, \bibinfo{person}{Li Li}, \bibinfo{person}{Liu
  Zhe}, \bibinfo{person}{Zhiming Liu}, \bibinfo{person}{Jacques Klein}, {and}
  \bibinfo{person}{Tegawendé F.~Bissyandé}.} \bibinfo{year}{2022}\natexlab{}.
\newblock \showarticletitle{Crex: Predicting patch correctness in automated
  repair of C programs through transfer learning of execution semantics}.
\newblock \bibinfo{journal}{\emph{Information and Software Technology}}
  \bibinfo{volume}{107043} (\bibinfo{year}{2022}).
\newblock
\urldef\tempurl%
\url{https://doi.org/10.1016/j.infsof.2022.107043}
\showDOI{\tempurl}


\bibitem[\protect\citeauthoryear{Yang and Yang}{Yang and Yang}{2020}]%
        {yang2020exploring}
\bibfield{author}{\bibinfo{person}{Bo Yang} {and} \bibinfo{person}{Jinqiu
  Yang}.} \bibinfo{year}{2020}\natexlab{}.
\newblock \showarticletitle{Exploring the Differences between Plausible and
  Correct Patches at Fine-Grained Level}. In
  \bibinfo{booktitle}{\emph{Proceedings of the 2nd International Workshop on
  Intelligent Bug Fixing}}. IEEE, \bibinfo{pages}{1--8}.
\newblock
\urldef\tempurl%
\url{https://doi.org/10.1109/IBF50092.2020.9034821}
\showDOI{\tempurl}


\bibitem[\protect\citeauthoryear{Yang, Lei, Mao, Lo, Xie, and Yan}{Yang
  et~al\mbox{.}}{2021a}]%
        {yang2021ground}
\bibfield{author}{\bibinfo{person}{Deheng Yang}, \bibinfo{person}{Yan Lei},
  \bibinfo{person}{Xiaoguang Mao}, \bibinfo{person}{David Lo},
  \bibinfo{person}{Huan Xie}, {and} \bibinfo{person}{Meng Yan}.}
  \bibinfo{year}{2021}\natexlab{a}.
\newblock \showarticletitle{Is the Ground Truth Really Accurate? Dataset
  Purification for Automated Program Repair}. In \bibinfo{booktitle}{\emph{2021
  IEEE 28th International Conference on Software Analysis, Evolution and
  Reengineering (SANER)}}. IEEE.
\newblock


\bibitem[\protect\citeauthoryear{Yang, Liu, Kim, Koyuncu, Kim, Tian, Lei, Mao,
  Klein, and Bissyand{\'e}}{Yang et~al\mbox{.}}{2021b}]%
        {yang2021were}
\bibfield{author}{\bibinfo{person}{Deheng Yang}, \bibinfo{person}{Kui Liu},
  \bibinfo{person}{Dongsun Kim}, \bibinfo{person}{Anil Koyuncu},
  \bibinfo{person}{Kisub Kim}, \bibinfo{person}{Haoye Tian},
  \bibinfo{person}{Yan Lei}, \bibinfo{person}{Xiaoguang Mao},
  \bibinfo{person}{Jacques Klein}, {and} \bibinfo{person}{Tegawend{\'e}~F
  Bissyand{\'e}}.} \bibinfo{year}{2021}\natexlab{b}.
\newblock \showarticletitle{Where were the repair ingredients for Defects4j
  bugs?}
\newblock \bibinfo{journal}{\emph{Empirical Software Engineering}}
  \bibinfo{volume}{26}, \bibinfo{number}{6} (\bibinfo{year}{2021}),
  \bibinfo{pages}{1--33}.
\newblock


\bibitem[\protect\citeauthoryear{Yang, Zhikhartsev, Liu, and Tan}{Yang
  et~al\mbox{.}}{2017}]%
        {yang2017better}
\bibfield{author}{\bibinfo{person}{Jinqiu Yang}, \bibinfo{person}{Alexey
  Zhikhartsev}, \bibinfo{person}{Yuefei Liu}, {and} \bibinfo{person}{Lin Tan}.}
  \bibinfo{year}{2017}\natexlab{}.
\newblock \showarticletitle{Better test cases for better automated program
  repair}. In \bibinfo{booktitle}{\emph{Proceedings of the 11th Joint Meeting
  on Foundations of Software Engineering}}. ACM, \bibinfo{pages}{831--841}.
\newblock
\urldef\tempurl%
\url{https://doi.org/10.1145/3106237.3106274}
\showDOI{\tempurl}


\bibitem[\protect\citeauthoryear{Ye, Gu, Martinez, Durieux, and Monperrus}{Ye
  et~al\mbox{.}}{2021}]%
        {ye2021automated}
\bibfield{author}{\bibinfo{person}{He Ye}, \bibinfo{person}{Jian Gu},
  \bibinfo{person}{Matias Martinez}, \bibinfo{person}{Thomas Durieux}, {and}
  \bibinfo{person}{Martin Monperrus}.} \bibinfo{year}{2021}\natexlab{}.
\newblock \showarticletitle{Automated classification of overfitting patches
  with statically extracted code features}.
\newblock \bibinfo{journal}{\emph{IEEE Transactions on Software Engineering}}
  (\bibinfo{year}{2021}).
\newblock


\bibitem[\protect\citeauthoryear{Ye, Martinez, Luo, Zhang, and Monperrus}{Ye
  et~al\mbox{.}}{2022b}]%
        {selfapr}
\bibfield{author}{\bibinfo{person}{He Ye}, \bibinfo{person}{Matias Martinez},
  \bibinfo{person}{Xiapu Luo}, \bibinfo{person}{Tao Zhang}, {and}
  \bibinfo{person}{Martin Monperrus}.} \bibinfo{year}{2022}\natexlab{b}.
\newblock \bibinfo{title}{SelfAPR: Self-supervised Program Repair with Test
  Execution Diagnostics}.
\newblock
\newblock
\urldef\tempurl%
\url{https://doi.org/10.48550/ARXIV.2203.12755}
\showDOI{\tempurl}


\bibitem[\protect\citeauthoryear{Ye, Martinez, and Monperrus}{Ye
  et~al\mbox{.}}{2019}]%
        {ye2019automated2}
\bibfield{author}{\bibinfo{person}{He Ye}, \bibinfo{person}{Matias Martinez},
  {and} \bibinfo{person}{Martin Monperrus}.} \bibinfo{year}{2019}\natexlab{}.
\newblock \showarticletitle{Automated Patch Assessment for Program Repair at
  Scale}.
\newblock \bibinfo{journal}{\emph{CoRR}}  \bibinfo{volume}{abs/1909.13694}
  (\bibinfo{year}{2019}).
\newblock
\urldef\tempurl%
\url{http://arxiv.org/abs/1909.13694}
\showURL{%
\tempurl}


\bibitem[\protect\citeauthoryear{Ye, Martinez, and Monperrus}{Ye
  et~al\mbox{.}}{2022a}]%
        {10.1145/3510003.3510222}
\bibfield{author}{\bibinfo{person}{He Ye}, \bibinfo{person}{Matias Martinez},
  {and} \bibinfo{person}{Martin Monperrus}.} \bibinfo{year}{2022}\natexlab{a}.
\newblock \showarticletitle{Neural Program Repair with Execution-Based
  Backpropagation}. In \bibinfo{booktitle}{\emph{Proceedings of the 44th
  International Conference on Software Engineering}} (Pittsburgh, Pennsylvania)
  \emph{(\bibinfo{series}{ICSE '22})}. \bibinfo{publisher}{Association for
  Computing Machinery}, \bibinfo{address}{New York, NY, USA},
  \bibinfo{pages}{1506–1518}.
\newblock
\showISBNx{9781450392211}
\urldef\tempurl%
\url{https://doi.org/10.1145/3510003.3510222}
\showDOI{\tempurl}


\bibitem[\protect\citeauthoryear{Yu, Martinez, Danglot, Durieux, and
  Monperrus}{Yu et~al\mbox{.}}{2019}]%
        {yu2019alleviating}
\bibfield{author}{\bibinfo{person}{Zhongxing Yu}, \bibinfo{person}{Matias
  Martinez}, \bibinfo{person}{Benjamin Danglot}, \bibinfo{person}{Thomas
  Durieux}, {and} \bibinfo{person}{Martin Monperrus}.}
  \bibinfo{year}{2019}\natexlab{}.
\newblock \showarticletitle{Alleviating patch overfitting with automatic test
  generation: a study of feasibility and effectiveness for the Nopol repair
  system}.
\newblock \bibinfo{journal}{\emph{Empirical Software Engineering}}
  \bibinfo{volume}{24}, \bibinfo{number}{1} (\bibinfo{year}{2019}),
  \bibinfo{pages}{33--67}.
\newblock
\urldef\tempurl%
\url{https://doi.org/10.1007/s10664-018-9619-4}
\showDOI{\tempurl}


\bibitem[\protect\citeauthoryear{Zhao, Li, Wang, Cai, Bissyand{\'e}, Klein, and
  Grundy}{Zhao et~al\mbox{.}}{2021}]%
        {zhao2021impact}
\bibfield{author}{\bibinfo{person}{Yanjie Zhao}, \bibinfo{person}{Li Li},
  \bibinfo{person}{Haoyu Wang}, \bibinfo{person}{Haipeng Cai},
  \bibinfo{person}{Tegawend{\'e}~F Bissyand{\'e}}, \bibinfo{person}{Jacques
  Klein}, {and} \bibinfo{person}{John Grundy}.}
  \bibinfo{year}{2021}\natexlab{}.
\newblock \showarticletitle{On the impact of sample duplication in
  machine-learning-based android malware detection}.
\newblock \bibinfo{journal}{\emph{ACM Transactions on Software Engineering and
  Methodology (TOSEM)}} \bibinfo{volume}{30}, \bibinfo{number}{3}
  (\bibinfo{year}{2021}), \bibinfo{pages}{1--38}.
\newblock


\bibitem[\protect\citeauthoryear{Zhao, Yu, Su, Liu, Zheng, Zhang, and
  Halfond}{Zhao et~al\mbox{.}}{2019}]%
        {zhao2019recdroid}
\bibfield{author}{\bibinfo{person}{Yu Zhao}, \bibinfo{person}{Tingting Yu},
  \bibinfo{person}{Ting Su}, \bibinfo{person}{Yang Liu}, \bibinfo{person}{Wei
  Zheng}, \bibinfo{person}{Jingzhi Zhang}, {and} \bibinfo{person}{William~GJ
  Halfond}.} \bibinfo{year}{2019}\natexlab{}.
\newblock \showarticletitle{Recdroid: automatically reproducing android
  application crashes from bug reports}. In \bibinfo{booktitle}{\emph{2019
  IEEE/ACM 41st International Conference on Software Engineering (ICSE)}}.
  IEEE, \bibinfo{pages}{128--139}.
\newblock


\end{thebibliography}

\end{document}